\documentclass[aps,twocolumn,amsmath,amssymb,secnumarabic,eqsecnum,pra]{revtex4-1}
\usepackage{graphicx}
\usepackage[colorlinks=false]{hyperref}

\newcommand {\be}{\begin{equation}}
\newcommand {\ee} {\end{equation}}
\newcommand {\bea}{\begin{eqnarray}}
\newcommand {\eea} {\end{eqnarray}}
\newcommand{\non}{\nonumber}
\newcommand{\bk}{{\bf k}}
\newcommand{\bx}{{\bf x}}

\newcommand{\cj}{{\cal J}}
\newcommand{\gj}{{\Gamma_\cj}}
\newcommand{\kj}{{\kappa_\cj}}
\newcommand{\rj}{{\rho_\cj}}
\newcommand{\nj}{{\nu(\cj)}}

\begin{document}
\title{Short-range Ising spin glasses: the metastate interpretation of replica symmetry breaking}
\author{N. Read}
\affiliation{Department of Physics, Yale University, P.O.\ Box 208120, New Haven, CT 06520-8120, USA}
\date{September 11, 2014}
\begin{abstract}
Parisi's formal replica-symmetry--breaking (RSB) scheme for mean-field spin glasses has long been
interpreted in terms of many pure states organized ultrametrically. However, the early version of this
interpretation, as applied to the short-range Edwards-Anderson model, runs into problems because as shown
by Newman and Stein (NS) it does not allow for chaotic size dependence, and predicts non-self-averaging
that cannot occur. NS proposed the concept of the metastate (a probability distribution over infinite-size
Gibbs states in a given sample that captures the effects of chaotic size dependence) and a non-standard
interpretation of the RSB results in which the metastate is non-trivial and is responsible for what
was called non-self-averaging. In this picture, each state drawn from the metastate has the
ultrametric properties of the old theory, but when the state is averaged using the metastate, the
resulting mixed state has little structure. This picture was constructed so as to agree both with
the earlier RSB results and with rigorous results. Here we use the effective field theory of RSB, in
conjunction with the rigorous definitions of pure states and the metastate in infinite-size systems,
to show that the non-standard picture follows directly from the RSB mean-field theory. In addition, the
metastate-averaged state possesses power-law correlations throughout the low temperature phase; the
corresponding exponent $\zeta$ takes the value $4$ according to the field theory in high dimensions $d$,
and describes the effective fractal dimension of clusters of spins. Further, the logarithm of the number
of pure states in the decomposition of the metastate-averaged state that can be distinguished if
only correlations in a window of size $W$ can be observed is of order $W^{d-\zeta}$. These results extend
the non-standard picture quantitatively; we show that arguments against this scenario are inconclusive.
More generally, in terms of Parisi's function $q(x)$,  if $q(0)\neq \int_0^1 dx\,q(x)$, then the metastate
is nontrivial. In an Appendix, we also prove rigorously that the metastate-averaged state of the
Sherrington-Kirkpatrick model is a uniform distribution on all spin configurations at all temperatures.
\end{abstract}
\maketitle

\section{Introduction}
\label{intro}

The problem of characterizing the equilibrium properties of classical spin glasses has been perplexing
for over forty years. The model that is still viewed as the theoretical paradigm is the
Edwards-Anderson (EA) model \cite{ea}, in which Ising spins interact with their nearest neighbors
by exchange interactions that are random and drawn independently, all from the same distribution;
the thermodynamic properties and equilibrium correlation functions must be found for
given values of these random variables, and so are themselves random quantities. Thus the EA Hamiltonian is
\be
H= -\sum_{(i,j)}J_{ij}s_is_j,
\label{EAham}
\ee
where $s_i=\pm 1$ is the Ising spin at site $i$. The sites are located at the positions $\bx_i$ which
lie on the vertices of a $d$-dimensional hypercubic lattice, with lattice spacing $1$, so $\bx_i\in
{\bf Z}^d$. For a finite-size version, only a portion $\Lambda$ of ${\bf Z}^d$ is used, with some
boundary conditions. The sum is over the edges $(i,j)=(j,i)$ of the graph ${\bf Z}^d$ (or $\Lambda$),
which connect vertices that
are nearest neighbors (under the Euclidean metric). Finally, the $J_{ij}$'s are
independent Gaussian random variables with mean zero and variance $J_0^2$, and an average over all
$J_{ij}$s will be written as the bracket $[\cdots]$. A magnetic field can be included by adding the
term $-h\sum_i s_i$ to $H$; we usually set $h$ to zero, or to a small value that tends to zero with
increasing system size (the latter is called an ``ordering field''). There are many review articles
and books on the spin glass problem; for a sample, see Refs.\
\cite{by,mpv_book,fh_book,ns_rev,ddg_book,bovier_book,mm_book,cg_book,ns_book,pan_book}.

Unlike the ferromagnetic Ising model, in which all $J_{ij}$ are positive, a random ``instance''
(or ``realization'') of the EA model typically has about half of its bonds $J_{ij}$ negative (here we
will assume that $\Lambda$ contains a finite number $N$ of vertices; the problem of taking the
thermodynamic limit will occupy us later). It is then not obvious what the ground-state (lowest-energy)
configuration will be: in general there is no spin configuration (i.e.\ a choice
of $s_i=\pm 1$ for all $i$) that makes $J_{ij}s_is_j$ positive simultaneously for all edges $(i,j)$.
This is referred to as ``frustration''. It is a property the EA model shares with many other examples
in statistical physics, and with many optimization problems. Frustration makes the task of finding the
ground state of a given instance with $N$ spins difficult, and in fact ``hard'', in the following sense
(from theoretical computer science \cite{moo_mer_book}, which will be used in this paragraph only).
In three dimensions or more, finding the ground state of a (finite-size) instance of the EA model
is ``NP-hard'', which means that in the worst case instance (here we forget the probability distribution
on the instances temporarily), the decision version---that is, to answer the yes/no question of whether
or not the ground state energy is less than some threshold---is NP-complete \cite{bara}, and so the
problem of actually finding the ground state is at least as hard as any decision problem in
the complexity class NP, or ``NP-hard'' (though not significantly harder \cite{moo_mer_book}). Hence
it is believed that the worst case requires a computation time that grows faster than any polynomial
in $N$, and may be exponentially large. (For planar two-dimensional examples, polynomial-time algorithms
are known.) The situation is not as clear if we consider typical instances of the EA as it was originally
defined (i.e.\ with a probability distribution of bonds), but there is little reason to expect a
polynomial-time algorithm there either. Heuristically, the reason for the difficulty of NP-hard
optimization problems, even when using randomized methods such as Monte Carlo, is that almost all
instances are frustrated, and then typically there exist many very different configurations that are
local optima and are good candidates for being the ground state (or global optimum). Similarly the
dynamics of the physical systems may (when the temperature $T$ is low) become stuck for long intervals
of time near
a local minimum that is not the ground state. Hence spin glasses are hard to equilibrate and exhibit
extremely slow relaxation to thermal equilibrium in the low-temperature region. The EA spin glass thus
serves as a paradigm for this kind of ``glassy'' behavior in many fields. Some of the references include
background about experiments \cite{by,fh_book,ns_rev,ns_book}, and some include reviews of the application
of spin-glass concepts to other fields, such as optimization and neural networks
\cite{mpv_book,fh_book,mm_book,ns_book}.


We will tackle the key problems of EA spin glasses that are central from the physical
perspective, and which are connected with the difficulties described in the previous paragraph.
EA argued that there is a thermodynamic phase transition in the infinite size limit of
their model at a nonzero transition temperature $T_c$ for zero magnetic field (provided the dimension
$d$ of space is high enough). Below the transition
temperature, the individual spins are ordered (in the sense that each one is found still to have
the same value after an arbitrarily large time has elapsed), but their signs are equally likely to be
positive or negative. We will use the basic concept of (classical) equilibrium statistical
mechanics, that a ``state'' of a system is identified with a probability distribution on the set of
configurations of the dynamical variables; a ``thermal'' average of a function of the variables over a
state in general will be denoted by $\langle\cdots\rangle$. The local magnetization $\langle s_i\rangle$
is the thermal average of the spin $s_i$ at ${\bf x}_i$ in some state with given random $J_{ij}$'s.
Then in the EA type of ordered or ``pure'' state (the precise definition of the concept of pure state
will be discussed extensively as we go on) the magnetization per site vanishes when averaged over
positions: $\lim_{N\to\infty}\sum_{i=1}^N \langle s_i\rangle/N=0$. However, the EA order parameter
$q_{\rm EA}$ which can be defined by the $N\to\infty$ limit of
\be
\sum_{i=1}^N \langle s_i\rangle^2/N
\label{overlap}
\ee
is non-vanishing below $T_c$ in such a state. As we will discuss (and similarly to the case of energy
minima discussed
above), there might be many distinct such ordered/pure states---each partially characterized by a set
$\{\langle s_i\rangle:i\in\Lambda\}$---that are not just related to one another by flipping (reversing)
the signs of all the spins $s_i$ as would be the case for the pure states in a conventional ferromagnet.
The basic issue
then for equilibrium statistical mechanics is to understand the phase structure of the low-temperature
phase in the infinite size limit. (It is generally believed that a transition at non-zero temperature
occurs only for $d>2$; that is the regime we discuss here.) Are there many pure states? How are they
characterized? Or is there
instead only slowing-down behavior, with only one or two thermodynamically-relevant pure states? Is there
a thermodynamic phase transition in a non-zero magnetic field, or only a crossover? How do the answers
to these questions depend on the dimension $d$ of space? These topics have been the subject of controversy
and there are two main schools of thought: the first is based on the replica-symmetry breaking (RSB)
mean-field theory \cite{par79} which suggests that there are many pure states and a transition in non-zero
magnetic field, while the second adheres to the scaling/droplet theory \cite{bm1,macm,fh} which assumes
that there are only two flip-related pure states below $T_c$, or in a magnetic field one pure state and
no transition. The discussion has continued to the present day; some recent entries and responses are
Refs.\ \cite{mb,pt,ykm,bill,middleton,bill2}.

In this paper, rather than proposing direct arguments for or against the existence of many states,
we will aim to determine what exactly the RSB theory says about the EA model in the infinite-size limit,
in terms of rigorously defined concepts for that limit, including the ``meta\-state'', which is a
probability distribution over thermodynamic states in a given sample, introduced by Newman and Stein (NS)
(and reviewed in Ref.\ \cite{ns_rev}). NS argued that the simplest form of the metastate that is
consistent both with results of the RSB mean-field theory and with rigorous results is what they called
the ``non-standard'' picture \cite{ns_rev}; it has a rich structure and is also consistent with many
numerical results. We argue in this paper that the ``RSB metastate'' that emerges by applying the methods
of RSB theory itself directly to study the meta\-state is exactly this ``non-standard'' picture. We
critique arguments (also reviewed in Ref.\ \cite{ns_rev}) that sought to show that this picture was
ruled out. We further characterize the RSB metastate quantitatively in terms of an exponent $\zeta$,
show that $\zeta=4$ in the RSB mean-field theory valid above $d=6$ dimensions, and speculate on its
behavior in lower dimensions. We propose that numerical work be done to investigate the value of $\zeta$,
in particular for low $d$. We hope that those on different sides of the controversy can agree about what
follows directly from the RSB approach, and we believe that such agreement might itself lead to further
progress in the field.

Let us summarize the results of the paper in slightly more detail, leaving full technical details
for later. The notion of the metastate was introduced by NS \cite{ns96b} to handle
the issue of possible ``chaotic size dependence'' (CSD) in spin glasses \cite{ns92}. That is, it may not
be possible simply to take the
limit of systems of increasing size in fixed samples, that is by increasing the size of $\Lambda$ by
retaining the random bonds in the existing system, and adding spins and bonds $J_{ij}$ (chosen
independently) around the exterior, because the state of the system (as determined by spin correlations
in any fixed region) may not approach a unique limit. The idea of the NS metastate is to handle this
by defining the probability of each such state as the frequency with which it occurs asymptotically as the
system size is increased without bound (there is a question of whether the limit of this probability
distribution itself exists).

Alternatively, averages over the metastate can be defined [following Aizenman
and Wehr (AW)] by taking an average over the bonds in a large outer region (much larger than, and
surrounding, the region in which the correlations are taken) in the limit in which the outer region goes
off to infinity \cite{aw}; this AW metastate has been argued \cite{ns97} to be the same as the NS
metastate. As usual for states in equilibrium statistical mechanics, a state drawn from the metastate may
not be a pure state, but could instead be a mixture of pure states. The concept of ``pure state'' that
we will use is only defined in infinite size, where it makes sense rigorously (even for disordered
systems that are not translationally invariant).

It follows that one way to understand the metastate is that it gives us information about how the state
of a system of given finite size, as seen in a subregion, depends on the bonds far away, when the
bonds nearby are given.
The dependence of the state on far-away single bonds, or finite sets of bonds, will be negligible in the
limit of large size. But it could still depend on the large set consisting of all the bonds far outside
the subregion, and that is described by the AW metastate. If such dependence occurs, it illuminates
the difficulty of determining the ground state, for example; if the ground state were independent of
far-away bonds, then it could be computed using only local information, a much easier task. For similar
reasons, the state in a subregion could depend on the sample size chosen, even though all bonds
are given (as described by the NS metastate).

We use the AW definition for an average over the metastate, and
apply the replica method; some of our results from this also appeared in Refs.\ \cite{par_comm,mari}, though
full details were not given, and they were not used to characterize the metastate explicitly. (NS responded
to Parisi's comment \cite{par_comm} in Ref.\ \cite{ns_resp}.) We show that the resulting metastate has the
structure of the non-standard picture of NS, as follows. First, each state drawn from it possesses the
properties of RSB, with a hierarchical organization into a countable number of pure states, and thus
essentially, the properties of the RSB solution hold (in some approximate sense) in any fixed
finite-size system. Second, at the same time, the metastate is ``dispersed'' (or ``non-trivial''),
that is, it is not supported on a single state. Some features of the RSB scheme, such as so-called
non-self-averaging, turn out to be due to this dispersal: a metastate-average shows no such effect,
as it must according to rigorous results \cite{ns96a,ns96b}.

Third, if one takes the average of the
state with respect to the metastate, what results is a mixed state. This can again be analyzed into
pure states but does not show the hierarchical structure. Instead, the disorder average of the
self-overlap as in eq.\ (\ref{overlap}) [where now the sum is over a finite subregion containing $N$ spins]
\be
\sum_{i=1}^N [\langle s_i\rangle^2]/N,
\label{mean_overlap}
\ee
and the correlation function
\be
C({\bf x}_i-{\bf x}_j)=[\langle s_i s_j\rangle^2],
\ee
tend to zero as $N$ (resp.\ $|{\bf x}_i-{\bf x}_j|$) $\to\infty$. Here the thermal average
$\langle\cdots\rangle$ is with respect to the metastate-averaged state (MAS), with given random $J_{ij}$'s
at all finite---i.e.\ not infinite---distances from the origin, and the disorder average $[\cdots ]$ is
over all these $J_{ij}$'s. This is true even at zero temperature, and means that the self-overlap as in
eq.\ (\ref{overlap}) [for the sum is over a finite subregion containing $N$ spins] in this MAS tends to
zero as $N\to\infty$, even if an ordering field is used. Because the self-overlap is non-zero in any
pure state in the MAS when $T<T_c$, this decay must be due to the presence of many such pure states in
the MAS, which typically have small overlaps with one another. Thus there is strong dependence
of the spin state in a region on the totality of bonds far away (though not on individual such bonds),
or on the system size: the metastate is {\em very} dispersed, and the pure states in the MAS are
uncountable, as they must be \cite{ns07}. (The authors of Ref.\ \cite{mari} had similar results to these
but stated that they
differed from the ``non-standard'' picture, and insisted on using an ill-defined notion of ``finite-size
pure states'', which was criticized in Ref.\ \cite{ns02}.)

We extend this NS non-standard picture here quantitatively by calculating $C({\bf x})$ using results from
the RSB approach. We find that
\be
C({\bf x})\sim |{\bf x}|^{-(d-\zeta)}
\ee
as $|{\bf x}|\to\infty$, where $\zeta$ is the
exponent mentioned already, which is $d$-dependent and universal, should have the same value both at zero
and non-zero temperature (below $T_c$), and $\zeta=4$ within the RSB scheme neglecting loop corrections.
(The result $\zeta=4$ follows from calculations in Ref.\ \cite{dkt,ddg_book}, where however it was not
applied to the metastate. In non-zero magnetic field, $\zeta$ is replaced by $\zeta'=3$
\cite{dkt,ddg_book}.)
If we sum $C(\bx_i-\bx_j)$ over $i$ ($j$) for which ${\bf x}_i$ (${\bf x}_j$) lie in a subregion $\Lambda_W$
of size $W\gg 1$ and use the pure-state decomposition of the MAS together with the clustering property
\be
\langle s_is_j\rangle_\alpha \to \langle s_i\rangle_\alpha\langle s_j\rangle_\alpha,
\label{clust}
\ee
which holds in the large distance limit for thermal averages $\langle\cdots\rangle_\alpha$ in a pure state
$\alpha$, we obtain
\be
\left[\sum_{\alpha,\beta}\mu_\alpha \mu_\beta\left(\sum_{i\in \Lambda_W}\langle s_i\rangle_\alpha \langle
s_i\rangle_\beta\right)^2\right]\sim W^{d+\zeta}
\ee
as $W\to\infty$ (up to a constant factor). (For simplicity we have written this expression as if the
decomposition is into a discrete set of pure states $\alpha$, though in fact that turns out not to be the
case; $\mu_\alpha$ is the weight of pure state $\alpha$ in the decomposition of the MAS, with $\sum_\alpha
\mu_\alpha=1$.)
Then for the overlap for the window, normalized ``per spin'' (that is, that divided by $N=W^d$), the
average of the square scales as $\sim W^{-(d-\zeta)}$. Examination of higher
moments in a similar way shows that they factorize into products as for a Gaussian distribution. We view
the result as describing the non-trivial metastate, and as holding for any typical $J_{ij}$s; then we can
say that the pure states in the MAS have a Gaussian distribution of per-spin overlaps. (As $W\to\infty$,
this becomes a $\delta$-function, as stated in Refs.\ \cite{par_comm,mari}.)

As the per-spin
self-overlap in each pure state is of order one, and treating mutual per-spin overlaps of two typical
pure states as a sum of statistically-independent variables, this result can be interpreted
as a central limit theorem. For zero temperature, each pure state is simply a discrete spin configuration,
and there is at most a finite number that can
be distinguished within any finite region. Then we can view the result as saying that the logarithm of
the number of pure states that can be distinguished within the window scales as $W^{d-\zeta}$. There is
a lower bound, $\zeta\geq 1$, that results from the short range of the interactions, because the influence
of distant bonds must be mediated by the spins on
the boundary of the region, so the logarithm of the number of pure states cannot be larger than
$W^{d-1}$; this is expected to hold at non-zero temperature also. The fact that $\zeta$ is larger than
$1$ means that the
states are somewhat robust against the effects of distant bonds; spin configurations must be influenced by
nearby bonds to some extent. If $\zeta=d$, then RSB itself will be absent, and we suggest that this occurs
at and below the lower-critical dimension of the EA spin glass, in which cases there is no transition
at positive temperature. $\zeta<d$ implies that the number of states in the MAS is uncountable, as stated
above.

It is natural to ask what is the analog of these results in the infinite-range Sherrington-Kirkpatrick (SK)
model \cite{sk}. In the SK model, for the (infinite-size) MAS based on the AW metastate defined in a
similar way as before, the power-law correlation above cannot hold because there is no notion of
distance that can be used. We show that in its place, in the MAS of the SK model, distinct spins are
simply uncorrelated, and in fact the joint distribution of spins in the MAS gives equal probability to
all spin configurations at any temperature $T\geq 0$ (this extends a result by NS \cite{ns03}); thus CSD is
pronounced.

The structure of the remainder of this paper is as follows: In Sec.\ \ref{sec:rsb_thy} we describe briefly
the broad context for the present work, including: replica symmetry breaking and its early interpretation;
other approaches, including the TAP equations, the cavity method, and the scaling/droplet theory; and
CSD and the strongly-disordered model. In Sec.\ \ref{sec:states_rsb}, we begin the main work of the paper
by first explaining the essential background, namely the rigorous definitions of Gibbs states and pure
states in infinite size systems, the metastate (in both AW and NS definitions), and the various scenarios
for the metastate. Then we introduce some ideas about fractal aspects of the metastate, and finally we use
the RSB scheme to determine the AW metastate of the RSB theory. Because the metastate that emerges from
this analysis agrees with the so-called non-standard picture, in Sec.\ \ref{sec:critique} we provide a
critique of arguments that sought to rule out this scenario for the metastate of the EA spin glass.
The paper closes with some discussion in Sec.\ \ref{sec:disc} of further topics: what may happen in lower
dimensions; the issue of stiffness (or rigidity) in the spin-glass phase; the application of the
metastate to finite-size systems and the question of global, as opposed to window, overlaps.
In Appendix \ref{sec:aw-sk} we discuss the analogs of both the AW metastate
and the MAS in the SK model, including a theorem that exactly describes the SK MAS at all temperatures as
a uniform distribution over spin configurations.

\section{Replica symmetry breaking and other theories}
\label{sec:rsb_thy}

Before we describe the results of this paper in more detail, in this section we will provide more
background and context. Most topics described here will reappear in the discussion later in the paper.
Readers who are experts on spin glasses, or those who wish to reach the results quickly, can skip to
the following section. In Sec.\ \ref{sec:rsb_early}, we describe the early history of the SK model,
the replica method, RSB theory, and early work on its interpretation, that were based on the idea
of  overlap distributions and the existence of many pure states or ``valleys''. In Sec.\ \ref{sec:cav},
we mention an alternative approach now known as the cavity method, that produces the results of RSB
without using replicas. In Sec.\ \ref{sec:issues}, we discuss some issues with the early interpretation.
In Sec.\ \ref{sec:scal/drop}, we introduce the leading alternative theoretical scenario to RSB, the
scaling/droplet picture. Finally, in Sec.\ \ref{sec:csd-sdm} we explain the idea of CSD with
particular reference to the strongly-disordered spin glass model of NS. The discussion of the metastate
and the various scenarios for it is central to this paper, and so appears not here but in the following
Section \ref{sec:states_rsb}.

\subsection{Replica symmetry breaking and early work on its interpretation}
\label{sec:rsb_early}

After the EA paper, SK \cite{sk} formulated the infinite-range model
which became the standard playground for spin-glass theorists for many years. In this model, there are
$N$ sites, and there is a bond $J_{ij}$ for every pair of sites $i$, $j$ (thus the underlying graph is
the complete graph on $N$ vertices), and the variance $J_0^2$ has to be replaced by $J_0^2/N$ to ensure
the existence of a finite thermodynamic limit, $N\to\infty$ with $J_0$ fixed, for the expected energy
(for example). In such a model, a mean-field approximation, formulated using EA's ``replica method'',
is expected to be
exact. But SK's proposed solution, while exhibiting a thermodynamic phase transition from a
high-temperature phase to a low-temperature spin glass phase (as in EA's theory), also had the
defect of having negative entropy at low temperature. In fact, it was soon shown that their solution,
called ``replica-symmetric'' because it maintains the permutation symmetry of the $n$ replicas or
copies in the limit as $n\to0$ (which is required in the replica method), was unstable to small
perturbations \cite{at}, and hence the true
saddle-point or mean-field solution had to involve somehow breaking the replica symmetry.

Shortly after, another attempt to solve the SK (or similar) model(s) was made by Thouless,
Anderson, and Palmer (TAP) \cite{tap}, by using a version of the Bethe-Peierls method adapted to a
spin glass, which produces a system of local self-consistency equations for the set of $m_i$ for all
$i$; nominally, $m_i=\langle s_i\rangle$ is the thermal average of the spin at $i$. Here the angle bracket
$\langle\cdots\rangle$ which stands for a thermal average means more precisely the average using the
normalized Gibbs weight which {\em for a system with $N$ finite} is defined by
$e^{-H/T}/Z$, where
\be
Z=\sum_{\{s_i=\pm 1\}} e^{-H/T}
\ee
is the partition function. $H$, $Z$, and $m_i$
are random as they depend on the $J_{ij}s$. However, while the average $\langle s_i\rangle$ is
clearly unique in a finite system, the solutions to the TAP equations turn out not to be unique below
$T_c$ (in fact, there are exponentially many \cite{te,dggo,bm_tap}), and the meaning of the $m_i$'s in a
solution must be understood in a more subtle fashion; this is essentially a form of the central
issues to be discussed in this paper.

Then, in seminal papers, Parisi \cite{par79} found a solution to the replica
mean-field equations of the SK model, that breaks the replica symmetry in an elaborate hierarchical
pattern, and the order parameter for the spin glass was found to be a function $q(x)$ on the
interval $x\in[0,1]$. For the SK model, it was found that $q(x)$ is non-negative, continuous,
monotonically increasing, constant over the half-open interval $(x_1,1]$ for some $x_1\in (0,1]$,
and $q(0)=0$ in the case of zero magnetic field.
(Actually, some later calculations were found to require summation over all the symmetry-related
inequivalent saddle points found by Parisi, which restores replica symmetry in a sense.) Even today, it
is not known how to make mathematical sense of this procedure. Parisi's replica-symmetry breaking (RSB)
solution to the SK model was shown not to violate physical requirements (for example, the entropy and
the reciprocal spin-glass susceptibility are non-negative for all $T\geq0$ \cite{par79}),
and to be marginally stable against perturbations \cite{dk}. By now it is known
that this solution is correct rigorously for the thermodynamics (the mean free energy) of
the SK model as $N\to\infty$ \cite{guerra,tala}.

After Parisi's work it was urgent to understand the meaning of the apparently successful RSB solution
of the SK model. At that time it became evident not only that there are distinct local minima (in
the sense that the energy increases if any moderate number of spins is flipped) of the SK Hamiltonian,
but also that for at least some pairs of minima, the ``barrier'' between the two minima increases with
system size $N$, leading to slow low-temperature dynamics, in line with the remarks in Sec.\
\ref{intro} \cite{somp,my}.
Here the barrier can be defined as the minimum (with respect to choice of a path) value of the maximum
(with respect to spin configurations on a given path) energy of any path of spin configurations
connecting two given energy minima. If the space of spin configurations can be partitioned into regions
or ``valleys'' separated by high barriers, then it is natural to interpret, for a
given sample, each solution
of the TAP equations (at least, those that are stable against small perturbations) with an average
taken within a ``valley'' surrounding an energy minimum. Of course, it is difficult to find a precise
definition of this idea in a finite-size system; for a careful attempt, see the final
chapter of Ref.\ \cite{mm_book}. (It may also be necessary to frame it in terms of free energies instead of
energy for the positive temperature case.)

Initially, there were some equilibrium quantities for which an expression in the RSB scheme was not
obvious. This was clarified by De Dominicis and Young \cite{dy}, who also suggested
a way to calculate such quantities using the TAP solutions, as an average over solutions, weighted
using the free energy $f_\alpha$ of each solution that follows from the TAP formalism. They further
rederived the replicated SK problem from the TAP equations, using replicas and some assumptions. Parisi
\cite{par83} used related ideas, though not the TAP equations, to interpret the Parisi function $q(x)$
in terms of a probability density for the overlap of two ``pure states'', with which the TAP solutions can
presumably be identified (again, we defer precise discussion of pure states until later; here we only
intend to describe the results in the language used by their authors). If $\langle\cdots\rangle_\alpha$
denotes an average over a ``pure state'' (or ``valley'') $\alpha$, then the (normalized) overlap of
two pure states was defined as
\be
q_{\alpha\beta}=\frac{1}{N}\sum_i\langle s_i\rangle_\alpha\langle s_i\rangle_\beta.
\ee
$q_{\alpha\beta}$ can be considered as kind of (inverse) distance between two pure states; when it is
large, the pure states are close together. For a TAP solution, $\langle s_i\rangle_\alpha=m_i$. Each
pure state has non-negative weight $w_\alpha\propto e^{-f_\alpha/T}$ ($\sum_\alpha w_\alpha=1$), so
\be
\langle\cdots\rangle = \sum_\alpha w_\alpha \langle\cdots\rangle_\alpha
\ee
gives the decomposition of the thermal probability measure as a ``mixture'' (a convex combination) of
pure states. Then Parisi defined the disorder-average distribution of overlaps
\be
P(q)=\left[\sum_{\alpha,\beta} w_\alpha w_\beta \delta(q-q_{\alpha\beta})\right],
\label{overlapdist}
\ee
and showed that $P(q)=dx/dq$, where $x(q)$ is the inverse of the function $q(x)$, which exists
because $q(x)$ in the RSB theory is monotonically increasing. Thus,
$x$ is the cumulative probability for $q$. Here it is presumed
that there is a ordering field, and that then each overlap $q_{\alpha\beta}$ is non-negative,
as is $q(x)$ in the RSB solution. From all this it follows that, for example, the average overlap defined
by equilibrium statistical mechanics
\be
q\equiv \left[\langle s_i\rangle^2\right]
\ee
(clearly averaging over positions makes no difference once the disorder average is performed)
is given by \cite{dy}
\be
q=\int_0^1 dx\,q(x),
\ee
or by $q=\int_0^1 dq\,P(q) q$ \cite{par83}, while the (disorder-averaged) Edwards-Anderson order
parameter \cite{ea} is a spin autocorrelation in the long time limit, and so should be understood as
\be
q_{\rm EA}=\left[\sum_\alpha w_\alpha \langle s_i\rangle_\alpha^2\right].
\ee
This is equal to $q(1)$ for the following reasons \cite{somp,dy}. In the RSB solution, $q(x)$ is found
to have a plateau, or interval of $x$ that includes $x=1$, on which it is equal to $q(1)$. As the
self-overlap $q_{\alpha\alpha}$ would be expected to be greater than or equal to $q_{\alpha\beta}$ for
$\beta\neq\alpha$, it is argued that the $\delta$ function in $P(q)$ is due to the self-overlaps of the
pure states, and also that these are independent of the pure state $\alpha$ and of the realization of
disorder, as $N\to\infty$. Then $q_{\rm EA}=q_{\alpha\alpha}=q(1)$ is self-averaging, and because
$[\sum_\alpha w_\alpha^2]$ (the length of the plateau) is non-zero, $w_\alpha$ is non-zero, so there
is a discrete set of pure states $\alpha$ that each have non-zero thermal weight, as has been
tacitly assumed in the notation like $\sum_\alpha\cdots$. At $T>T_c$, the situation becomes trivial
as there is a single pure state, and in zero magnetic field $q(x)=0$ for all $x$. (In a non-zero
magnetic field $h$, there is a similar situation, with a non-trivial $q(x)$ function below the
de Almeida-Thouless line in the $h$-$T$ plane \cite{at}. We will continue to focus on $h=0$.)

Further developments followed this interpretation. These included the result that the overlap
distribution $P_\cj(q)$ (defined as in eq.\ (\ref{overlapdist}), but with the average over disorder
omitted) does not approach the averaged $P(q)$, even as the system size tends to infinity; this was
referred to as ``non-self-averaging'' \cite{ybm,mpstv}. Others were the ultrametric organization
of the pure states, where the distance between pure states was defined using the overlaps
$q_{\alpha\beta}$ \cite{mpstv,mv}, and the fact that the free energies $f_\alpha$ of the pure states,
which are random, are independent and exponentially distributed \cite{mpv1,dt}. There are later rigorous
results that are consistent with aspects of this picture, including stochastic
stability \cite{ac}, the Ghirlanda-Guerra identities \cite{gg}, and a proof of ultrametricity \cite{panch}.
These results are mostly for the SK model, but there are also some for the EA model, for example,
the recent Ref.\ \cite{cms}.


\subsection{Cavity method}
\label{sec:cav}

In view of the probabilistic interpretation of the RSB theory of the SK model \cite{dy,par83}, and of
the results mentioned that connect the approach using the TAP equations with the RSB solution \cite{dy},
it was logical to look for an approach that could give the results of the RSB theory without using any
$n\to0$ limits. Such an approach was found in Ref.\ \cite{mpv} (see also the book, Ref.\ \cite{mpv_book}),
and is now called the cavity method. In this approach, one considers the effect of adding one more spin
to an SK model of $N$ spins, using knowledge of the solution for $N$ to obtain that for $N+1$ spins.
The resulting equations that characterize the state of the added spin can then be applied
``self-consistently'' to all the spins, as they are equivalent after the disorder average. Further,
the solution should approach a stable fixed distribution that is reproduced on adding one spin, up to
some rescalings. [The variance of the bonds must be changed from $J_0^2/N$ to $J_0^2/(N+1)$
when adding one spin, and this is taken into account in the formulas.] The cavity method can be viewed
as a development of the TAP equations (which were based on the Bethe-Peierls method), and leads to
related equations. It has been developed further and connected with methods termed ``belief propagation''
and ``survey propagation'' which can be applied to more general optimization (and other) problems (for
a review, see \cite{mm_book}).

In order to include RSB at the ``one-step'' level (which is needed below $T_c$ at zero magnetic field,
or below the de Almeida-Thouless line more generally), it is necessary in the cavity method to assume
that the $N$ spin system has a countably-infinite set of so-called pure states labeled $\alpha$, each
with a
free energy $f_\alpha$. As above, the meaning of these pure states or ``valleys'' in finite size ($N$
spins) is not entirely clear. If they can be defined, then the number of such pure states in a finite
size should be finite; replacing the number by infinity is an approximation that, for the goal of
deriving the limiting distributions as $N\to\infty$, may be innocuous.
At the next level of RSB, these pure states are organized into clusters, and this is repeated
hierarchically infinitely many times to produce the full RSB scheme. (For details, see Refs.\
\cite{mpv,mpv_book}.) The results of the cavity method are then found to be fully consistent with those
of the proposed interpretation of the RSB solution of the SK model, given the assumptions we have mentioned.

\subsection{Issues with the early interpretation}
\label{sec:issues}

A key property that distinguishes pure states from mixtures of pure states under {\em rigorous}
definitions of these terms (which we discuss later), and was used by Parisi \cite{par83}, is the
clustering property (\ref{clust}). This raises a problem with the interpretation in Ref.\ \cite{par83}
because it requires that the system size be infinite, and then one takes the limit of infinite separation
of the sites in such a system. This makes sense for the well-defined notion of a pure state in a short-range
model in infinite Euclidean space (such as the EA model), which we will explain in detail later. Leaving
aside the question of the infinite-size limit for a moment, in the finite-size SK model there is
no useful notion of distance, as all vertices are nearest neighbors, and so clustering cannot be defined
in the usual way. (However, replacements for the notion of clustering in an
infinite-range model have been proposed \cite{by,mpv_book}.) This is just one of the problems with using
a notion of pure states defined in a finite-size system.

In any case, Parisi's conclusions were supposed to hold also for the EA model, in which pure states (under
a rigorous definition) should exist. But there is a further difficulty: the results derived by Parisi
\cite{par83} are in fact for averages over disorder in a finite system, and the infinite-size limit is
taken after the average (as always in the study of the SK model). However the language used in the paper
and others from the same era suggests that the statements are for properties of (say) pure states in an
{\em infinite} system. The problem is that the existence of a limit for an average, or even for a
distribution function for quantities that are random because they depend on the random $J_{ij}$s, does
not necessarily mean that there exists a limiting object (even a random one) whose properties one has
obtained. For example, it was stated that the barriers between valleys go to infinity in the limit, but
it is not clear that there is any limit at all for the pure states (``valleys'') that were supposedly
defined in a finite system. In general, the object in question (the Gibbs state, in this example) could
vary from size to size, and still for any one large size have the distribution that was obtained. In
the literature this issue---CSD---apparently went unnoticed until Ref.\ \cite{ns92}. Instead, the existence
of limiting objects in a sample with given $J_{ij}$s, such as a collection of pure states organized
hierarchically, was tacitly assumed. This picture, in which the existence of the limit is supposed to be
unproblematic, gives what NS call the ``standard SK picture'' of the states in the EA spin glass. It has
been conclusively ruled out for both the SK \cite{ns92} and the EA models \cite{ns96a}, and is no longer
defended by Parisi \cite{mari}.

A separate issue concerns the use of overlaps which, following the SK model, were defined as the sum
of overlaps of all the spins, divided by the number $N$, in a finite system. It has been argued that
such global overlaps can be misleading as to the pure-state structure of a system \cite{fh1} (these
authors also argued that many pure states would not occur in the EA model \cite{fh2}). This problem can
be avoided by using overlaps in a finite subregion or ``window'' only, and taking a limit as the window
size goes to infinity.

\subsection{Scaling/droplet theory}
\label{sec:scal/drop}

In the interpretation of the RSB theory a central ingredient is the existence of many pure states. A
leading alternative to this scenario is the scaling/droplet theory \cite{bm1,macm,fh}. This approach
began with simple scaling ideas about the effect of a flip in boundary condition from nominally periodic
to nominally antiperiodic, which produces a ground state with a domain wall relative to the original
ground state. The change in energy is random, with a distribution whose width was argued to scale as a
power of the length of the wall, and the wall has some fractal dimension. As acknowledged at the beginning
of the second of Refs.\ \cite{fh}, these ideas do not in themselves imply that there is a single pair
of ground states, or a single pair of pure states at positive temperature, in zero magnetic field below
$T_c$ in sufficiently high dimension $d$ (or just one in each case at non-zero magnetic field). (Again,
the explanation of the sense in which there could be more than one requires the use of the thermodynamic
limit, or infinite system size, and will be discussed later.) However, in practice, the scaling/droplet
theory refers to an approach that {\em assumes} there is only a single such pair ({\it et cetera}), and
then shows using ideas of droplets as low-energy excitations that the low-temperature behavior could
nonetheless be non-trivial \cite{fh}. It is noteworthy that according to this theory there is neither a
true (thermodynamic) transition in non-zero magnetic field, nor CSD (asymptotically
at large sizes) for zero magnetic field below $T_c$ when, say, periodic boundary conditions are used.


\subsection{CSD and strongly-disordered model of a spin glass}
\label{sec:csd-sdm}

The notion of CSD \cite{ns92} has been described informally already. It has not been proved
that CSD actually occurs for the EA model (with spin-flip-invariant boundary conditions) in any dimension
$d$. However, for the SK model, there
is a direct proof by NS \cite{ns03} that {\em any} spin configuration is the ground state for some---in
fact, infinitely many---sizes $N$ (to be precise, the description of these ground states uses the idea of a
window, which is defined in the following section). This result is reviewed briefly in Appendix
\ref{sec:aw-sk}. That the ground state changes chaotically with size could have been anticipated from the
(somewhat heuristic) derivation of the RSB results by the cavity method \cite{mpv}. There the free energies
$f_\alpha$ associated to each pure state change each time a spin is added (as do the spin expectation
values $m_i$), and do so randomly, though the distribution remains a product of exponentials. From this
one would expect the free energies of different pure states to sometimes cross, so that at zero
temperature the ground state will sometimes jump from one configuration to another, consistent with Ref.\
\cite{ns03}.

There is a short-range model in finite dimensions $d$ in which CSD can be established for some $d$:
the strongly-disordered spin-glass model \cite{ns94,bcm}). One may think of this model as defined in
finite size in the limit in which the distribution of bonds $J_{ij}$ becomes very broad, rather than
Gaussian (so disorder is strong). The condition for the width of the distribution to be sufficiently
large will increase with system size, making the model physically unrealistic, but it is nonetheless of
some theoretical interest. We focus on finding the ground state. Consider the bonds in order of decreasing
magnitude. In the limit, the first one, say that for the edge $(i_0,j_0)$,  will be very large, much larger
than all other bonds ending on $i_0$ or $j_0$. Then there is no doubt that the sign of $s_{i_0}s_{j_0}$ in
the ground state will be ${\rm sgn}\,J_{i_0j_0}$. With this fixed, one can then find the next strongest
bond (in magnitude), and repeat the procedure. Each edge found to be the next strongest in magnitude will
be ``accepted'', meaning that it determines the relative signs of the two spins at its ends, unless those
spins are already connected by a path of previously-accepted bonds, in which case one can ignore that bond
and proceed to the next strongest in turn. With free boundary conditions, this procedure, which is a
greedy algorithm, will eventually find the absolute ground state, up to an overall choice of sign of all
the spins, due to the global spin-flip symmetry. Moreover, the set of edges accepted forms a ``maximum
spanning tree'', a set of edges of the graph that touches every vertex and has no cycles (our graphs are
always assumed connected), and the procedure is Kruskal's greedy algorithm for finding it. (This ``maximum
spanning tree'' problem is equivalent to the ``minimum spanning tree'' problem by a simple change of sign
of the weights on the edges from $|J_{ij}|$ to $-|J_{ij}|$.) If instead of free boundary conditions we use
fixed boundary conditions, with all the spins on the boundary of the hypercube fixed to values $\pm 1$ with
independent probabilities $1/2$ for each spin (independent of the bonds), then the procedure is similar.
The vertices on the boundary can be viewed as already connected by very strong bonds (in the``maximum''
version of the algorithm), or ``wired''. Once a spin in the interior is connected to the fixed spins on
the boundary by a path of accepted edges, its value is determined. In this case, the interior of the
hypercube is eventually occupied by a set, or forest, of trees, not a single tree, and the trees in
the forest span all the vertices; this is then called a maximum (or in the other version, minimum)
spanning forest.

According to the analysis of minimum spanning trees on lattices in $d$ dimensions carried out by NS
\cite{ns94} (as corrected in Ref.\ \cite{jr}; the central result of that work has recently been extended
in Ref.\ \cite{addario}), if one looks in a subregion (``window'') of size $W$, as the
system size $L\to\infty$, then for $d>6$ one will see pieces of the total tree, each of which is a tree
with one or more roots on the boundary of the window. There are of order $W^{d-6}$ pieces of size a given
fraction (say, 1/2) of $W$ in diameter; each of these contains of order $W^6$ vertices. For $d<6$, only
a single large piece will be visible in any given window. These results also hold for the maximum spanning
forest produced using the wired boundary condition; the conditions far from the window make little
difference to the pieces visible in the window.

Then for $d>6$, the strongly-disordered spin-glass model displays both sensitivity to the fixed-spin
boundary condition, and CSD \cite{ns94}. For the former, notice that for any one piece (tree) in
the window,
the choice between a set of spin values and the set in which all are flipped is determined by the single
spin on the boundary that the piece eventually makes contact with (outside the window). Most likely,
the flips for distinct large pieces in the window are determined by distinct boundary spins, though
occasionally, and especially for small pieces, there may be a connection with another large piece that
occurs outside the window, in which case they must flip together when the boundary spin is flipped.
Then we conclude that the number of distinct ground states visible in the window, as the boundary spins run
over their set of $2^{{\cal O}(L^{d-1})}$ distinct values, will be $2^{{\cal O}(W^{d-6})}$.

To understand CSD, first we note that if one then considers such models of increasing sizes $L$, constructed
by adding additional spins and bonds at the boundary, keeping the interior bonds the same, then the
local structure of the spanning tree near the origin converges to a limit (see Ref.\ \cite{ns94} and
references therein). However, even
for free boundary conditions, for which the trees visible in a fixed size window all become connected
somewhere outside the window, these connections occur further and further from the window as $L$ increases.
Hence each tree-piece in the window in the spin-glass model will again flip randomly as the size
increases. Thus for $d>6$ the model exhibits CSD, and again the number of ground
states that can be distinguished in the window will be of order $2^{{\cal O}(W^{d-6})}$.


\section{Gibbs states, metastates, and RSB}
\label{sec:states_rsb}

In this section we make more precise definitions of the important concepts, and present the central
arguments of this paper. We begin with the general concepts of Gibbs states and pure Gibbs states, both
defined for infinite-size systems. Then we use those concepts in defining the NS concept of the metastate,
and describe the main alternative scenarios for the metastate structure of the EA model. We explain how
correlations in the MAS can be used to characterize the number of pure states occurring in the metastate
in relation to ideas drawn from fractal-cluster models of a spin glass. Then we reach the main calculation
of the paper, the use of RSB in a finite-dimensional effective field theory to examine the metastate; it
is found to agree with the so-called non-standard scenario, and the exponent $\zeta$ is obtained. Finally,
we summarize the results and some scenarios that arise.

\subsection{Gibbs states and pure states}
\label{sec:gibbs}

We begin with some notation; our conventions usually follow the papers of NS. We consider the lattice
with vertices ${\bf Z}^d$, with bonds $J_{ij}$ connecting nearest neighbors. We write $\cj=\{J_{ij}\}$
for a set of $J_{ij}$s as $(i,j)$ ranges over edges of the infinite lattice. It is convenient to view the
set of all these random bonds as already given, even though for finite size only a finite number of them
enter. We write $\nj$ for the probability distribution of $\cj$, which is always a product over $(i,j)$ of
a Gaussian for each $J_{ij}$. Likewise we write $S=\{s_i\}$ for the spin values $s_i$ as $i$ ranges over
the lattice sites. We refine the notation $\Lambda$ for a finite portion of the lattice, as follows: we
will make use of subsets of
the lattice that are hypercubes $\Lambda_L$, defined for odd $L$ as containing $L^d$ vertices and centered
at the origin, with faces perpendicular to the coordinate axes. We will frequently refer to triples of
nested hypercubes, $\Lambda_L$, $\Lambda_R$, and $\Lambda_W$. We always suppose that the sizes are in
the sequence
\be
W\ll R\ll L,
\ee
and well separated as indicated here. The uses of these different hypercubes will be explained as we go on.

At a given temperature $T$, we define the Gibbs state for a finite system on $\Lambda_L$ as the probability
distribution on the set of $S$ defined by the probabilities $e^{-H_{L,\cj}/T}/Z$, where $H_{L,\cj}$ is
the EA
Hamiltonian (\ref{EAham}) for $\Lambda_L$; the magnetic field is usually set to zero. We denote this
(random) Gibbs state as $\Gamma_{\cj,L}$; it depends on $\cj$ through $H_{L,\cj}$. $H_{L,\cj}$ and
hence also $\Gamma_{\cj,L}$ are only fully determined when a boundary condition (b.c.)
has been specified. Possible choices include (i) free b.c.s, for which there are terms in
$H_{L,\cj}$ only for edges $(i,j)$ that lie within $\Lambda_L$; (ii) periodic b.c.s, in
which there are in addition edges that connect vertices on each face to their counterpart on the opposite
face; for those bonds connecting the faces perpendicular to the $\mu$th coordinate axis ($\mu=1$, \ldots,
$d$), we can use $J_{ij}$ for the edge $(i,j)$ with $x_{i\mu}=(L-1)/2$, $|x_{i\mu'}|\leq (L-1)/2$
($\mu'\neq\mu$), and $x_{j\mu}=(L-1)/2+1$, $x_{j\mu'}=x_{i\mu'}$ ($\mu'\neq\mu$); (iii) fixed-spin
b.c.s, in which the spins on the surface of $\Lambda_L$ are fixed to specified values.
The b.c.\ chosen can depend on $L$, or can be random, but in all cases, acceptable
b.c.s should be chosen independent of the bonds within (i.e.\ in the interior of) $\Lambda_L$.
For fixed-spin b.c.s, CSD can occur even in a non-disordered Ising ferromagnet, if the boundary spins
change with system size for all sizes (the sensitivity of the ground state in this model to random 
boundary conditions in any fixed size was noted in Ref.\ \cite{vanenter}); a similar phenomenon can occur 
in a spin glass even within the scaling/droplet theory \cite{ns92}. As this is of less interest, we 
usually assume (except when stated) free or periodic b.cs, which do not break spin flip symmetry (this 
assumption was implicit earlier also).

In an infinite system, such as our lattice ${\bf Z}^d$, the previous definition of a Gibbs state
determined by a Hamiltonian $H$ can no longer be used whenever (as in our case) $H$ involves an infinite
sum of terms that does not converge. Instead there is a well-established definition for the notion of a
Gibbs state $\Gamma$ in a system that can be infinite \cite{ruelle_book,sinai_book,georgii,bovier_book}
(it is also
termed a Dobrushin-Lanford-Ruelle, or DLR, state). It goes as follows (for a succinct account, see
Chapter 1 of Ref.\ \cite{ruelle_book}): consider
the conditional probability distribution for any {\em finite} set of spins, conditioned on all the spins
not in the finite set. The ratios of these conditional probabilities for any two spin configurations $S$,
$S'$ (where the spins not in the finite set are fixed to the same values) are proportional to
$e^{-[H(S)-H(S')]/T}$. This is unambiguous provided the interactions in $H$ are sufficiently short-range
so that the difference $H(S)-H(S')$ (constructed by term-wise subtraction) is a finite
(or more generally, a convergent) sum; this condition is clearly satisfied in the EA Hamiltonian with
nearest-neighbor interactions. In a finite system, this definition of a Gibbs state reduces to the usual
finite-size definition as given above. In our EA model with given $\cj$, we will denote infinite-size
Gibbs states generically by $\Gamma_\cj$.

Given a Hamiltonian $H$ and a temperature $T$, the definition allows a Gibbs state to be patched together
from finite-spin probabilities. In a finite system, the Gibbs state is unique (as above). But in an
infinite-size system this patching allows for different
b.c.s to be built in ``at infinity'' in general. Hence there may be many distinct Gibbs
states for the given Hamiltonian and temperature. The set of Gibbs
states is convex; that is, given two Gibbs states $\Gamma_1$, $\Gamma_2$, any ``convex combination''
(which is to be viewed in terms of the values of the probabilities for each spin configuration in these
states) $\lambda \Gamma_1 +(1-\lambda)\Gamma_2$ ($0\leq \lambda\leq 1$) is also a Gibbs state. The set of
all Gibbs states is also compact, and hence there are ``extremal'' Gibbs states that are not
convex combinations of any other Gibbs states. These extremal Gibbs states are also called ``pure'' states,
while non-extremal Gibbs states are termed ``mixed''. Then any Gibbs state can be decomposed uniquely into
a convex combination of pure states; as the latter may
be uncountable, in general the decomposition may take the form of an integral over the pure states with
some measure \cite{ruelle_book,sinai_book,georgii}. Moreover, the pure states can also be equivalently
characterized among the Gibbs states in general in other ways, for example clustering of {\em all}
correlations holds in an infinite-size Gibbs state if and only if it is pure
\cite{ruelle_book,georgii,bovier_book}.

It is worthwhile to mention the case when $T=0$, even though it is simply a special case of the preceding
definitions. At zero temperature a ground state is a Gibbs state. Here a ground state is defined to be
a single spin configuration $S$ that has the property that if any finite number of spins are flipped,
the energy stays the same or increases. (In our EA models with continuous distributions of bonds and
infinite size, the probability of a ground state for which the energy stays the same under flipping
some finite set of spins is zero, and the inequality becomes strict. We assume this from here on.)
In a finite system, there can be only one ground state (except for the two-fold degeneracy that is a
consequence of spin-flip symmetry in zero magnetic field in our models), but in an infinite system,
the definition allows the existence of many ground states. A ground state (in infinite size) is clearly
a pure (extremal) state. In infinite size systems, arbitrary convex combinations of ground states are
again Gibbs states, though not supported on single spin configurations; except in the case of a convex
combination of symmetry-related ground states, such as those that are related by a flip of all spins when
the magnetic field in the Hamiltonian is zero, these non-extremal zero-temperature Gibbs states are not
often discussed.

The preceding definition for pure states is the ``official'' one that we will use in this paper. It only
comes into play when the system is infinite; there is no evident idea of how to make a corresponding
precise definition in a finite system, in which there is a unique Gibbs state. (A finite-size Gibbs state
can be decomposed as a
convex combination in many ways. The procedure of first partitioning the configuration
space into a union of disjoint subsets, and then defining a conditional Gibbs distribution on each subset
as the analog of a pure state,
as in the ``valleys'' idea, is not the most general decomposition. Even if this procedure is used, the
choice of partition is not unique, and the decomposition into so-called pure states remains at best
fuzzy.)

It should be emphasized that the definitions of Gibbs and pure Gibbs states remain valid
if the Hamiltonian is not translationally invariant, for example if it involves quenched disorder as in
the EA model (this is clearly stated at the outset of Chapter 1 in Ref.\ \cite{ruelle_book}).
There have been repeated claims in the literature that this definition of pure states is somehow not
suitable for systems that are not translationally invariant, especially those like spin glasses
\cite{mari,mm_book}; sometimes these claims appear to be based on the attempt to interpret RSB using
the approach of Ref.\ \cite{par83}, based on the SK model and on overlaps defined using the set of all
spins in a finite-size system. These claims have been criticized by NS; see especially Ref.\ \cite{ns02}
and references therein. They have instead used overlaps, defined in finite windows, of the ``official''
pure states in an infinite system, as follows.

Given two pure states $\alpha$, $\beta$, we can compute overlaps $q_{\alpha\beta}$ in the window
$\Lambda_W$, with the conventional ``per spin'' normalization:
\be
q_{\alpha\beta}=\frac{1}{W^d}\sum_{i\in \Lambda_W}\langle s_i\rangle_\alpha\langle
s_i\rangle_\beta.
\label{overlaps}
\ee
We note that it is not obvious that $q_{\alpha\beta}$ has a limit as
$W\to\infty$, or that it is independent of the position of the window. We return to this issue in
Sec.\ \ref{sec:rigid} below; it is not important presently. For a given Gibbs state $\gj$ (which in the
EA model depends on the bonds $\cj$), we define the overlap distribution by
\be
P_{\cj,\gj}(q) =\sum_{\alpha,\beta}w_{\gj,\alpha}w_{\gj,\beta} \delta(q-q_{\alpha\beta}),
\label{eq:ovdist}
\ee
where for simplicity we assumed that the decomposition of $\gj$ into pure states,
\be
\langle\cdots\rangle_\gj=\sum_\alpha w_{\gj,\alpha}\langle\cdots\rangle_\alpha
\label{eq:puredecomp}
\ee
is discrete; $w_{\gj,\alpha}\geq0$ are the weights in the decomposition, with $\sum_\alpha
w_{\gj,\alpha}=1$.
This overlap distribution depends on the choice of window, and on the state $\gj$ from which it was
constructed. One of the central points of this paper is to argue that these window overlap distributions
should be used in calculations with the replica method.

\subsection{CSD and metastates}
\label{sec:csd}

The general definition of Gibbs states $\Gamma$ permits the existence of many such states at the given
temperature $T$, and the question then is how to obtain one that is physically relevant. A natural
procedure would be to try to take the limit of finite systems (with specified b.c.s),
in which the Gibbs state is unique. In systems with translationally-invariant Hamiltonians, such as the
Ising ferromagnet, the limiting Gibbs state indeed exists, and its decomposition into pure states forms
the basis for a discussion of its physical properties (in particular, correlation functions).

In systems with quenched disorder, such as the EA model, the finite size Gibbs state $\Gamma_{\cj,L}$
depends on $\cj$. If a limiting Gibbs state $\Gamma_\cj$ exists, it too depends on $\cj$. We should
emphasize here that such a state is equivalent to the complete set of correlation functions of
finitely-many spins, or equivalently to the marginal probability distribution for the spin configuration
in any finite region (for example, $\Lambda_W$) conditioned on the full set of bonds $\cj$. The existence
of the limit means
by definition that for any fixed finite set of spins, the marginal probability distribution for that
set obtained from $\Gamma_{\cj,L}$ (for large enough $L$, so that the set is a subset of $\Lambda_L$)
tends as $L\to\infty$ to that obtained from $\Gamma_\cj$. In the simplest case, one would have existence
of the limit for almost all $\cj$ [under the probability distribution $\nu(\cj)$ for $\cj$]. In fact, it
can be proved generally that either for almost all $\cj$ the limit exists, or for almost all $\cj$ it
does not exist \cite{ns92}. The latter alternative means that for almost all $\cj$ there is CSD.

The idea of a metastate (so-named by NS \cite{ns96b}) is that, more generally than a limiting Gibbs
state, one may
have in the limit a probability distribution over infinite-size Gibbs states $\Gamma_\cj$. Just as a state
is probability distribution on spin configurations, a metastate is a probability distribution on states.
The measure for the distribution is conditional on $\cj$ and is denoted $\kappa_\cj$; we sometimes write
$\kappa_\cj(\Gamma)$ for its formal density in the space of $\Gamma=\Gamma_\cj$'s (formal because we do
not wish to enter into the question of how to integrate over the infinite-dimensional probability
distributions $\Gamma$). (In the special case when there is a limiting Gibbs state $\Gamma_\cj$,
$\kappa_\cj(\Gamma)$ will be a $\delta$-function on that $\Gamma_\cj$.) The existence of the limiting
distribution (metastate) is itself not automatic, and might not occur if the size dependence is extremely
chaotic.

There are actually two definitions of metastates in the literature. The earlier one is due (somewhat
implicitly) to AW \cite{aw}. Their insight was that as the boundary (consisting of
the faces of $\Lambda_L$) is pushed off to infinity, the disorder near the boundary is always further
away than all the bonds at fixed distance from the origin that constitute $\cj$ in the limit. Thus there
may be disorder at infinity in the limiting Gibbs state
that is not fixed by conditioning on $\cj$; the resulting probability distribution for Gibbs states
$\Gamma_\cj$ is the AW metastate. Formally, their metastate can be defined using the following procedure.
Consider the Gibbs state $\Gamma_{\cj,L}$ in the hypercube $\Lambda_L$ (for some b.c.s).
The hypercube can be viewed as containing an ``inner'' region $\Lambda_R$, and the remainder is the
``outer'' region; for the bonds, the $J_{ij}$ for edges $(i,j)$ in $\Lambda_R$ are in the inner region, and
the remainder, those with one or both ends not in $\Lambda_R$, are in the outer region. We can consider
the probability distribution on states $\Gamma_{\cj,L}$ due to the bonds in the outer region, conditioned
on those in the inner region. Then take the limit $L\to\infty$ first, followed by $R\to\infty$ (or one
might use $L\to\infty$ with $L=2R+1$, or similar; the precise way of taking the limit ought not to
be important). If this limit exists, we obtain the AW metastate. That is, it exists if the probability
distribution over states does; we recall that a state is itself equivalent to the set of marginal
``thermal'' probability distributions for hypercubes $\Lambda_W$ for given $\cj$, for all $W$.

A second definition is due to NS \cite{ns96b}. In their construction, there is no separation into inner
and outer regions. We consider the sequence of sizes $L=L_0$, $L_0+2$, $L_0+4$, \ldots, for some initial
size $L_0$ which is arbitrary; a choice of random $\cj$ is given.
In each size we have the Gibbs state $\Gamma_{\cj,L}$. For a given ``window'' $\Lambda_W$, we obtain from
each $\Gamma_{\cj,L}$ a corresponding marginal probability distribution on the spins in $\Lambda_W$ (it
consists of $2^{W^d}$ real numbers, with the single constraint that their sum is unity). Using the first
$n$ members of the sequence of $L$'s, we can obtain an empirical distribution on the  $2^{W^d}$-dimensional
space of probability distributions on the spins, which is a $\delta$-function with weight $1/n$ on each
point that occurs in the sequence of the first $n$ sizes. Then in the limit as $n\to\infty$, it may be
that this empirical distribution tends to a limit. If this is the case for every window $\Lambda_W$
(i.e.\ as $W\to\infty$ at the end), then the resulting distribution on states $\Gamma_\cj$ is called
the NS metastate.

In both constructions we have emphasized that the limits taken may not exist. However, there are ideas
of compactness that can be used for the sequences of probability distributions. For
an infinite sequence of points in a compact space, there must be limit (or accumulation) points, which are
approached arbitrarily closely, and infinitely many times, by the sequence. Likewise, for sequences
of distributions, there can be ``relative compactness'', and then there will be limit distributions
\cite{billingsley1,billingsley2}. The existence of limit points/distributions can also be described by
saying that the sequences have convergent subsequences. In this case, limiting distributions exist;
the question is whether the limit is unique. For the AW metastate, it has been shown that the subsequence
limits exist \cite{aw,ns97}, though uniqueness has not been proved. The NS metastate can also be shown
to exist for some subsequence of such a subsequence, and to equal the corresponding AW metastate
\cite{ns97}.

There are some results on other models that illuminate the relation between the
two approaches \cite{kulske,bg} (see also Appendix \ref{sec:aw-sk} for a discussion of the AW metastate
of the SK model). It appears that the AW metastate may be the simpler and more robust procedure, in the
sense that it may exist in the limit (i.e.\ uniquely) in cases in which the NS metastate does not.
Suppose that for given bonds in the inner
region, the probability distribution for the marginal of the states (defined in the window $\Lambda_W$)
due to disorder in the outer region approaches a limiting form as $L=2R+1\to\infty$, so that the AW
metastate exists. In the NS approach, in each size $L$ in this sequence we choose bonds in the outer
region, and find the state. For large $L$, the distribution for that $L$ approaches the AW metastate.
{\em If the states for each $L$ are independent}, then by the usual (weak) law of large numbers the
empirical distribution over the space of states will approach the AW metastate distribution, and so the
NS and AW metastates are the same. However, it could also be that the AW metastate exists, but that when
comparing systems of different sizes, they are not independent. If the correlations are weak enough,
the empirical distribution can still tend to AW metastate. But it can also happen that the limit of
the empirical distribution does not exist. For example, in the random-field Ising model \cite{kulske},
the disorder in the outer region essentially boils down to a sum of $N$ random variables (the random
fields), and because the distribution of this sum has width of order $N^{1/2}$ (and so is large), the
sign of this variable determines which of
two Gibbs states, say $+$ and $-$, is selected. (Here $N$ corresponds to $L^d$.) At any particular $L$,
this sign is equally likely to be $+1$ or $-1$, and the AW metastate exists. But in the empirical
distribution over sizes, the empirical probability for the $+$ state is the fraction of sizes for which
the sum is positive, and this is a random variable that does not approach a limiting value almost surely.
This occurs because the sizes are not independent; if the signs had been independently $+$ or $-$,
with probability $1/2$, for each $L$, the frequencies would have self-averaged to $1/2$. But in the
correlated case, the frequencies are random variables that do not self-average, though there is in turn
a well-defined distribution (the arcsine law) for the fraction of times the sum is $+$. (Nonetheless, if
one uses frequencies with which states occur among a sufficiently spread-out subsequence of sizes---such
that the states for each size used are drawn independently---then agreement with the AW metastate
is recovered, in accordance with the result stated above \cite{ns97}.) In this situation
in which the NS metastate does not exist, one can attempt to define the behavior of the whole random
sequence of states as $L\to\infty$ instead \cite{kulske,bg}. Now that these potential issues have been
noted, we will generally assume that both the AW and NS metastates do exist.

At zero temperature and zero magnetic field, the states $\Gamma_\cj$ that occur in the metastate for
free or periodic b.c.s will be
equal-weight superpositions of a ground (i.e.\ pure) state and its global spin flip, as a direct
consequence of the constructions using finite size systems, because in each finite size the unique
Gibbs state is a corresponding superposition of finite-size ground states (with probability one). The
pair obtained may vary chaotically with size, or with the realization of disorder at infinity
(corresponding to the two constructions of the metastate). Such behavior, and its direct analog at
nonzero temperature and zero magnetic field, in which each $\Gamma_\cj$ is a superposition of two
flip-related pure states, has been termed the ``chaotic pairs'' picture by NS \cite{ns96b}. At nonzero
magnetic field,
and also for b.c.s such as fixed spins or magnetic fields on the boundary,
the spin-flip symmetry is broken, and each pair of pure states is replaced by a single pure state.


The metastate of the NS strongly-disordered model of a spin glass (see Sec.\ \ref{sec:csd-sdm}) can be
readily obtained from the theory
in Ref.\ \cite{ns94}, with the updates in Ref.\ \cite{jr}. The case of fixed b.c.s is
the simplest to analyze \cite{ns97}. For $d>6$, each spin in $\Lambda_L$ is on a single tree that has a
root on the boundary, at which the value $\pm 1$ of the spin is fixed independently of the bonds in
the interior (and with probability $1/2$ for each value). The value of each spin on that tree is
determined by the spin on the boundary times the sign of the bonds on the unique path connecting the spin
to the boundary. As $L$ is made larger in a given sample $\cj$, the forest of trees in any finite
region $\Lambda_W$ converges to a limit, while the
boundary spin, to which each tree is attached, is chosen again independently for each $L$. Hence (as
discussed above in Sec.\ \ref{sec:csd-sdm}) all the
spins on each tree flip randomly with increasing size, or for fixed size, they all flip as the boundary
spin is changed. Each state $\Gamma_\cj$ in the fixed-spin b.c.\ metastate is supported
on a single spin configuration. Thus in a window $\Lambda_W$, where there are of order $W^{d-6}$ pieces
of size of order $W/2$, there are $2^{{\cal O}(W^{d-6})}$ distinct ground states that can be
distinguished, and each has equal probability. Note that in this model, both the AW and NS metastates
exist and are the same. For free b.c.s, the metastate is argued to be essentially the
same (except for the equal superposition of the global-flip-related ground states that occurs in each
$\Gamma_\cj$), though the details are less clear \cite{ns97}.

We will later use the MAS $\rj$, defined by
\be
\langle\cdots\rangle_\rj=[\langle\cdots\rangle_\gj]_\kj.
\label{rjavdef}
\ee
It can be shown to be a Gibbs state \cite{ns97}, so it has a decomposition into pure states also. Then we
can define a window overlap distribution $P_{\cj,\rj}(q)$ similarly as for any $\gj$.

\subsection{Scenarios for the metastate of the EA model}
\label{sec:scen}

We may now describe (following NS \cite{ns96b}) the leading scenarios for the EA spin glass in terms
of the metastate
for the low-temperature phase below $T_c$ (if there is a transition), for both zero and nonzero magnetic
field $h$. There are four logical classes of possibilities, because the metastate could be either trivial
(supported on a single Gibbs state) or nontrivial, and in either case a Gibbs state drawn from the
metastate could be either trivial (a pure state, or at strictly zero magnetic field, the equal-weight
superposition of a pair of flip-related pure states) or nontrivial (a mixture of trivial states).
Each of the four classes includes a range of possibilities that differ from each other in further
details. The theories that have been described earlier correspond to particular members of these classes.

At one extreme, there is the behavior assumed in the scaling/droplet model: the metastate is trivial; it is
supported on a single Gibbs state $\gj$ that is trivial. In this case there is no transition in a magnetic
field.

The other extreme includes scenarios inspired by RSB. One possibility is that the metastate
is supported on a single Gibbs state both for $h=0$ and $h\neq 0$, but the Gibbs state has a
non-trivial decomposition into pure states (at least for nonzero temperature). In particular, in what NS
\cite{ns96b} called the standard SK picture, the Gibbs state $\gj$ has the non-trivial properties arising
from Parisi's RSB scheme, including countably-many pure states organized hierarchically, and
non-self-averaging of the window overlap distribution $P_{\cj,\gj}(q)$ as $W\to\infty$. The latter
property, however, runs afoul of consequences of translation invariance
of the disorder distribution $\nj$ in the EA model \cite{ns96a,ns96b}, and so as an interpretation of
RSB theory the standard picture can be viewed as already excluded. (That may not be true for other
scenarios in this class, for example an uncountable decomposition into pure states \cite{ns96b}, provided
it exhibits self-averaging.)

The third class of possibilities is those in which the metastate $\kj$ is non-trivial, and so is each $\gj$
drawn from it. In particular, in a possible interpretation of RSB that NS termed the non-standard SK
picture, and called ``maximally mean-field-like'', each Gibbs state drawn from the metastate has
all the features of RSB (and so for $T>0$, $\gj$ is a non-trivial mixture of pure states both for $h=0$
and $h\neq0$). The argument based on translation invariance of $\nj$ implies that the overlap
distribution $P_{\cj,\rj}(q)$ of the MAS $\rj$, and also the
metastate average of the overlap distribution $[P_{\cj,\gj}(q)]_\kj$ in the Gibbs states $\gj$,
must self-average as $W\to\infty$, that is, be independent
of the bonds $\cj$ \cite{ns96a,ns96b}. Thus this means that, unlike in the standard picture, what was
called non-self-averaging is due to the non-trivial nature of the metastate. In fact, the set of pure
states occurring in the metastate, or alternatively in the MAS, must be uncountable \cite{ns07}.

Finally, there could be a non-trivial metastate, but each $\gj$ drawn from it could be trivial---the
chaotic pairs picture mentioned above. There is no issue about self-averaging for the chaotic pairs
picture, because the overlap distribution $P_{\cj,\gj}(q)$ of each $\gj$
is trivial, consisting for strictly zero magnetic field of a pair of $\delta$-functions at
$\pm q_{\rm EA}$ with equal weight (in a magnetic field, there is instead a single $\delta$-function
at $q=q_{\rm EA}$), and $q_{\rm EA}$ self-averages.

Some of the scenarios become equivalent at $T=0$. In all cases (note we continue to assume a continuous
distribution of bonds $J_{ij}$), at $T=0$ any Gibbs state $\gj$ drawn from the
metastate is trivial and its overlap distribution $P_{\cj,\gj}(q)$ has the same form as described for the
chaotic pairs scenario, for either zero or non-zero magnetic field, with $q_{\rm EA}=1$; hence one
might say the distribution self-averages. Nonetheless, some of the corresponding metastates still differ.
Both the scaling/droplet and standard SK picture assume/predict that at $T=0$ the metastate is supported
on a single trivial Gibbs state. Both the non-standard--SK and the chaotic-pairs pictures
assume/predict that at $T=0$ the metastate is supported on many trivial Gibbs states. In Ref.\ \cite{ns02}
it is asserted that the standard and non-standard pictures are equivalent
at $T=0$, for reasons that are not clear to us, but may be due to considering only the overlap
distributions.

Note that none of these statements about the metastate rule out the possibility that other pure/ground
states {\em exist} in the (infinite) system; they only describe what is obtained by analyzing the
metastate, and thus what is thermodynamically relevant in samples that are large and typical. In
particular, scenarios such as standard SK in which many pure states are present in the (single) Gibbs
state in the $T>0$ metastate presumably imply the existence of additional ground states also.

\subsection{Fractal aspects of metastates}
\label{sec:frac}

In the analysis of the NS strongly-disordered model of a spin glass in space dimension $d$,
fractals play a significant role at least for $d>6$. The picture of the model suggests more
general scenarios in which fractals are involved; we call these models fractal-cluster models
(similar ideas have been advanced in Ref.\ \cite{wf}), and we describe these for zero temperature first.
Thus suppose that the vertices of the lattice are partitioned once and for all (in a $\cj$-dependent
fashion) into infinite subsets (or clusters), each of which is a random fractal of the same dimension $D$,
so that
the number of vertices on one cluster that intersects a box $\Lambda_W$ is of order $W^{d-D}$ (at least
when the intersection has linear size $W/2$). Suppose there are ground states in which the spins on each
cluster have fixed relative sign with one another, but the choice of overall sign for each cluster is
arbitrary; it is ``controllable'' from infinity. Finally, a possible metastate is that in which each
choice of signs for all the clusters has approximately equal probability; for fixed-spin b.c.s, each state $\Gamma_\cj$ is a single spin configuration. Then there are of order
$2^{{\cal O}(W^{d-D})}$ ground states that can be distinguished from one another using the spin
configurations in $\Lambda_W$ only.

The value of the dimension $D$ in such a model can be obtained from spin correlation functions.
We write $\langle\cdots\rangle_{\Gamma_\cj}$ for the average in a Gibbs state $\Gamma_\cj$ (which here
is a single ground state, though the case of a uniform combination of two flip-related ground states
gives the same results for even correlation functions). We write $[\cdots]_{\kappa_\cj}$ for the average
over the $\gj$s with distribution given by the metastate $\kappa_\cj(\Gamma)$, and $[\cdots ]_{\nu(\cj)}$
for the average over $\cj$ with distribution $\nu(\cj)$. The correlations in a given $\Gamma_\cj$ drawn
from the metastate,
\be
\langle s_{i_1}\cdots s_{i_n}\rangle_\gj
\ee
will equal $\pm 1$ for all $n$ because $\Gamma_\cj$ here is a single spin configuration. Nonetheless, the
correlations depend on $\Gamma_\cj$, so that if an average over $\Gamma_\cj$ is performed using the
metastate, the result may be smaller than 1 in magnitude. Specifically, for the $n=2$-point function, if
the probability for each of the two flip-related states of each cluster is $1/2$ (independently), then
$[\langle s_i s_j\rangle_\cj]_\kj=\pm 1$ if $i$ and $j$ are on the same cluster, and $0$ if they are not;
in the first
case the sign is determined by the disorder and by the detailed behavior of the model. If we square
to eliminate this sign, and average over $\cj$, then because of the random ($\cj$-dependent) geometry of
the fractal we will obtain for all $k=1$, $2$, $3$, \ldots
\be
\left[\left[\langle s_i s_j\rangle_\gj \right]_\kj^{2k}\right]_{\nu(\cj)}\sim
\frac{1}{|\bx_i-\bx_j|^{d-\zeta}},
\ee
as $|\bx_i-\bx_j|\to\infty$, and the exponent $\zeta=D$, the fractal dimension of the clusters. We see
that the spin correlations in the $\cj$-dependent metastate-averaged state (MAS) $\rho_\cj$ [see eq.\
(\ref{rjavdef})]
can be non-trivial even at zero temperature. This yields information about the metastate. In
the strongly-disordered model, and in the present more general fractal-cluster models with equal probability
assigned to each flip of a cluster, the $2k$th moment is independent of $k$. This behavior indicates the
fractal-cluster nature of the underlying metastate.

Let us consider similar correlations using the metastate of a more general model, such as the original
EA model. We do not need to restrict ourselves to zero temperature. In the spin-glass ordered phase
(in zero magnetic field), correlations of an even number of spins such as $\langle s_i s_j\rangle_\gj$,
in a Gibbs state
$\gj$ drawn from the metastate, presumably tend to a (random) nonzero limit as $|\bx_i-\bx_j|\to\infty$.
At large separation, the situation is then similar to that at zero temperature. Then if the metastate
is sufficiently non-trivial, $[\langle s_i s_j\rangle_\cj]_\kj=\langle s_i s_j\rangle_\rj$
(and other even correlations in $\rj$) may decay with distance. The statistical properties of
these correlations should be universal, with exponents that are independent of temperature for $T<T_c$.
We would like to know the typical decay of the absolute value of the two point function $\langle s_i
s_j\rangle_\rj$, for example. This may be difficult to obtain, but the mean square, for which we
define notation
\be
C(\bx_i-\bx_j)=\left[\langle s_i
s_j\rangle_\rj^2\right]_{\nu(\cj)},
\label{cx}
\ee
is again a good starting point, and one can also consider the $2k$th moments as above.

We expect that when the metastate is non-trivial, we will have power-law behavior
\be
C(\bx_i-\bx_j)\sim \frac{1}{|\bx_i-\bx_j|^{d-\zeta}}
\ee
up to a constant factor, as $|\bx_i-\bx_j|\to\infty$ with a universal $T$-independent exponent $\zeta$.
Clearly $\zeta$ must obey
$\zeta\leq d$ in all cases. Actually, it is not clear if the metastate must be trivial when $\zeta=d$;
a decay slower than a power might be possible. We note the possibility that
in some cases, the full statistical behavior might not be described by this single exponent $\zeta$, but
a spectrum of exponents that generalize $\zeta$ to other moments might be needed, as for multifractals.

The question arises of whether or not such behavior must {\em always} be explained in terms of fractal
clusters. Let us return to the zero-temperature case for simplicity. In general, it might not be correct
to use the fractal cluster picture as above, even if the modification that the probabilities for the
two flip-related configurations of each cluster may not be equal is made. Suppose we compare two
ground states of a spin system. There will be vertices at which the spin configurations are the same,
and others at which they are opposite; we can color these black and white respectively. These two sets
partition space into regions we can call domains. One can form a system of ``domain walls'', defined as
surfaces on the dual lattice that locally are normal to each edge on which the relative sign of the spins
in the two states at the two ends is different. (The walls are well-defined, whereas an attempt to
partition the space into distinct domains of the two colors may not be, because domains of the same color
can meet at corners, and it is not obvious how to connect the regions.) In the fractal-cluster picture,
the edges on the domain walls for two ground states of the model are always a subset of the boundaries
separating the clusters; they run along some of the same surfaces, regardless of which two ground states
are considered. But in the case of the ground states of an EA model, if it has many ground states, the
domain walls might not always run through the same edges, and so there are no well-defined clusters.
But because we have defined the exponent $\zeta$ through the 2-point function $C(\bx)$, the idea now is
to use this to {\em define} an effective or average fractal dimension for the domains.

In the fractal-cluster models, the fractal dimension was used to characterize the number of pure
(i.e.\ ground) states distinguishable in a window $\Lambda_W$. We can do the same for the general models;
we start with the zero-temperature case. We use the marginal distribution of $\rj$ for a window
$\Lambda_W$; this is defined
as the average over the metastate of the marginal distribution of $\gj$ for the window. (Recall that
states $\gj$ were in fact defined by using such marginal distributions on a sequence of windows of size
tending to infinity.) The number of ground (pure) states visible in the window is the number of spin
configurations that have non-zero probability in the marginal distribution of $\rj$. (A more sophisticated
definition would be the exponential of the Shannon entropy of the marginal of $\rj$.) Because the
correlation function $C(\bx)$ decays with distance only because of the multiplicity of ground states,
and using the simplest model assumption that each one has approximately equal probability, the analogy with
the fractal-cluster models suggests that the logarithm of the number of ground states grows as
$W^{d-\zeta}$.

For non-zero temperature, we would like similarly to characterize the growth in the number of pure
states that can be distinguished in a window. The MAS $\rj$ is convenient for this purpose. Because
it is itself a Gibbs state \cite{ns97}, it can be decomposed as a convex sum/integral of (infinite-size)
pure Gibbs states. We denote averages in each pure state $\Gamma_{\cj,\alpha}$ for fixed $\cj$, indexed
by $\alpha$, as $\langle\cdots\rangle_\alpha$ (with $\cj$-dependence implicit as before), and the
decomposition is
\be
\langle\cdots\rangle_\rj=\int d\alpha\, \mu_{\cj,\alpha}\langle\cdots\rangle_\alpha,
\ee
where $\mu_{\cj,\alpha}$ is the density of the measure for $\alpha$, with $\int d\alpha\,
\mu_{\cj,\alpha}=1$ (we note that there will be an uncountable continuum of pure states involved). The
set of pure states $\Gamma_{\cj,\alpha}$ must be a subset of those that occur in the decomposition of all
the $\gj$ that occur in the metastate, but it is possible that there are pure states that do not occur
in $\rj$ with non-zero density in their neighborhood in $\alpha$. The pure states in $\rj$ form a
convenient set of ``accessible'' pure states that we can attempt to characterize.

To introduce the window, we wish to pass to marginal distributions. If we consider not $\Gamma_{\cj,\alpha}$
but its marginal for the window $\Lambda_W$, it may be that pure states with different index $\alpha$
become identical distributions on the window. If in the decomposition we pass to the marginal
distribution of each $\Gamma_{\cj,\alpha}$, and lump together those that become equal, we obtain a similar
decomposition over a reduced space, which is (a subspace of) the space of probability distributions on
the window:
\be
\langle\cdots\rangle_{\rj,(W)}=\int d\alpha_{(W)}\,
\mu_{\cj,\alpha,(W)}\langle\cdots\rangle_{\alpha,(W)},
\ee
where the subscript $(W)$ denotes passage to marginal distributions on the window---the average can
be applied to functions of the spins in $\Lambda_W$. In principle, we wish to find the entropy of
this ``marginal'' version $\mu_{\cj,\alpha,(W)}$ of $\mu_{\cj,\alpha}$. The space of probability
distributions on $\Lambda_W$ has dimension $2^{W^d}-1$. The distribution $\mu_{\cj,\alpha,(W)}$ on that
space may well be continuous. If we assume as we did at zero temperature that the measure
$\mu_{\cj,\alpha,(W)}$ is approximately uniform on its support, then instead of either counting the number
of points in its support, or finding its volume, we should probably find the dimension of the support,
which is at most $2^{W^d}-1$. By a similar picture as for zero temperature, because of the form of
$C(\bx)$, we will argue that this dimension in fact scales as $W^{d-\zeta}$ (at least under a somewhat
coarse-grained definition of the dimension).

We can also connect this picture of the pure states visible in a window in the MAS
with an overlap distribution. We use the preceding decomposition of the MAS, and the relevant
overlap distribution (for the overlap $q_{\alpha\beta}$ in the same window $\Lambda_W$) is defined by
\be
P_{\cj,\rj}(q) =\int\int d\alpha_{(W)}\, d\beta_{(W)}\,
\mu_{\cj,\alpha,(W)}
\mu_{\cj,\beta,(W)} \delta(q-q_{\alpha\beta}).
\ee
We will see later that this distribution has support at both positive
and negative $q$. For simplicity, let us focus on the average over $\cj$,
$\left[P_{\cj,\rj}\right]_{\nu(\cj)}$.

The width of the overlap distribution can be obtained by calculating its second moment. One expects in
fact that the mean of the overlap distribution is zero, so the second moment is the variance. (This
differs from
the case in the Parisi RSB scheme, but has been expected in the present and related contexts before
\cite{par_comm,mari,ns_rev}.) We can relate this second moment to the correlation function C(\bx). We use
the pure state decomposition of $\rj$ (squared). Then in each pure state, the long-distance behavior of
the 2-point function is
\be
\langle s_i s_j\rangle_\alpha\sim \langle s_i\rangle_\alpha\langle s_j\rangle_\alpha
\ee
as $|\bx_i-\bx_j|\to\infty$ by the clustering property which holds because we have genuine pure states.
If the window is large enough, then the asymptotic behavior is a good enough approximation
over most of the range of the position sum. Consequently, we find
\be
\int dq\,q^2 \left[P_{\cj,\rj}\right]_{\nu(\cj)} \sim W^{-2d} \sum_{i,j\in \Lambda_W} C(\bx_i-\bx_j)
\ee
as $W\to\infty$. Then the behavior of the correlation function immediately implies that
\be
\int dq\,q^2 \left[P_{\cj,\rj}\right]_{\nu(\cj)} \sim W^{-(d-\zeta)}
\ee
as $W\to\infty$. This confirms that the mean of $q$ is small or zero. This result immediately invites
comparison with the central limit theorem, at least if the distribution $[P_{\cj,\rj}]_{\nu(\cj)}$ is
Gaussian,
which we will discuss later. Thus, if we can think of the window overlap (not normalized per spin) of
one typical pure state with all the others as a sum of independent random variables with similar variances
of order $W^\zeta$ (as in the fractal-cluster models), the number of these variables must be of order
$W^{d-\zeta}$. Alternatively but similarly, one can think of a Gaussian distribution of the per-spin
overlaps of one with all the others, in a space of some unknown dimension; then the dimension must be
of order $W^{d-\zeta}$. We will now refine this picture.

The underlying basis for these statements that infer a dimension of a space from the width of an overlap
distribution can be visualized geometrically. We begin with the zero-temperature case. The spin
configurations in $\Lambda_W$ can be visualized as the vertices of a hypercube at $(s_1,s_2,\ldots,s_{W^d})$
in $W^d$-dimensional space (each $s_i=\pm 1$ as before). We imagine that all the ground states that can be
distinguished in $\Lambda_W$ are statistically similar. Then if the ground states are distributed
approximately uniformly over {\em all} vertices of the hypercube, the empirical distribution of per-spin
overlaps between pairs
of ground states will have most of its weight in an approximately Gaussian peak with mean zero and
variance of order $W^{-d}$. (The width is due to the geometry of the hypercube, and does not depend on
the number of ground states; we neglect fluctuations due to the dependence of the ground states on $\cj$,
assuming the number is sufficiently large.) Hence the larger variance
$\sim W^{-d+\zeta}$ must be due to correlations among the ground states. In the case of the fractal-cluster
model, the ground states are arranged on the vertices of a polyhedron which is a ``hyper-rhomboid'' of
dimension $\sim W^{d-\zeta}$ embedded in the hypercube above. It is a hyper-rhomboid because each
cluster that intersects $\Lambda_W$ can be flipped independently of all the others, and such a flip
corresponds to a set of parallel edges of the hyper-rhomboid. In general, that will not be the case,
but we can visualize the ground states as vertices of some polyhedron that would be formed by connected
each vertex (ground state) to some of its nearest neighbors, with distance measured as the (scaled)
Hamming distance,
\be
d_{\alpha\beta}= 1 - q_{\alpha\beta},
\ee
which is related to the overlap $q_{\alpha\beta}$ (defined as above) of two ground state configurations
$\alpha$, $\beta$. Again, the ground states (vertices of the polyhedron) should be distributed uniformly
over some region; a reasonable guess is that this region is topologically a hypersphere of some dimension.
This should at least hold over large Hamming distance $d$ (i.e.\ those of order 1), though at small
distances it may be that the hypersphere is thickened so that its dimension is larger on those scales.
The hypersphere is embedded into the hypercube of all spin configurations on $\Lambda_W$, and may
deviate from a geometrically perfect object to some extent, but we will model it as a hypersphere
nonetheless. Then the variance of the distribution of overlaps (or Hamming distances) will be $W$ to
the power of
minus the dimension of the hypersphere. The precise form of the hypersphere
will of course depend on $\cj$. Thus this picture captures the idea of ground-state-dependent domains, and
of the effective dimension of the set of ground states, which is the number of coordinates needed to
parametrize them, neglecting the small scale structure. Hence it shows that the dimension (on large scales)
of the hypersphere is of order $W^{d-\zeta}$, and the number of ground states is the exponential of this
quantity.

For the non-zero temperature case, we wish to have a similar picture. The dimension of the space of
(marginal) probability distributions on $W^d$ spins is $2^{W^d}-1$, and the distributions may vary
continuously as the pure states vary. However, for the purpose of calculating the overlap of two pure
states, their sets of spin expectations $\langle s_i\rangle_\alpha$ ($i\in \Lambda_W$), are
sufficient information. Such a set of numbers labels a point that now can be inside or on the surface of
the $W^d$-dimensional unit hypercube we had before, instead of only at its vertices as at $T=0$. Then
a simple picture to explain the overlap distribution is that the pure states again correspond to points
close to a roughly hyperspherical surface of dimension $\sim W^{d-\zeta}$. It is plausible that the
pure states (reduced to the window) themselves are well described by this number of parameters. We note
that for this case the Hamming distance should be replaced by
\be
d(\alpha,\beta)=\frac{q_{\alpha\alpha}+q_{\beta\beta}}{2}-q_{\alpha\beta},
\ee
which reduces to the Hamming distance when the self-overlaps $q_{\alpha\alpha}$ equal $1$.

Stronger bounds can be placed on the possible values of $\zeta$, at least at zero temperature.
In the fractal cluster model, each cluster
must be path-connected to infinity, otherwise one of the states obtained by flipping it would not be a
ground state. Hence $\zeta\geq 1$. More generally, one can argue that the logarithm of the number of
possible ground states in a window is at most of order $W^{d-1}$, as the ground state in the window
is uniquely determined if the spin values on the boundary of the window are given.
Put another way, the result that the width of the per-spin overlap distribution is of order $W^{\zeta-d}$,
with $\zeta\geq 1$, shows that some significant correlations are built into the pure states
in the metastate.


\subsection{The RSB metastate}
\label{sec:rsb_meta}

Now we will show how the metastate implied by the RSB theory can be calculated using the RSB scheme. Our
method involves an ansatz that is a slight extension of the Parisi ansatz \cite{par79}, and also resembles
techniques used to study magnetic-field, temperature, and disorder chaos, and finite-size fluctuations of
the free energy, within the RSB approach, see for example \cite{par83,somp_unpub,kondor,kv,am} (such
chaotic behavior also follows from the
scaling/droplet model \cite{bm87}). Several of the results appeared in Refs.\ \cite{par_comm,mari}, though
they were not explicitly described in terms of the metastate, and the method may have been different.

As the replica method has been discussed in many places, we will only describe it briefly here. Following
EA, one uses $n$ copies or ``replicas'' of the system, with the initial target of calculating the mean of
the free energy $F=-T\ln Z$ of the system, using the formal relation
\be
[\ln Z]=\lim_{n\to0}\frac{[Z^n]-1}{n}.
\ee
Here the system size should be finite so that $F$ and the partition function $Z$ make sense. For
finite-dimensional systems that are statistically homogeneous (with say periodic b.c.s),
we follow the approach of early authors, using a Hubbard-Stratonovich decoupling, and obtain an effective
field theory description for $[Z^n]$ as a functional integral over configurations of a real, symmetric,
matrix-valued field $Q_{ab}(\bx)$ ($Q_{ab}=0$ for $a= b$), where $a$, $b$ run from $1$ to $n$, and $n$
must be taken
to zero at the end. Here $Q_{ab}(\bx_i)$ has the meaning of a coarse-grained version of the product of
spin variables $s_{ia}s_{ib}$ in two copies $a$, $b$ of the system. Then we have
\be
[Z^n]=\int {\cal D}[Q] e^{-S},
\ee
where the effective action $S$ has a Landau-Ginzburg-Wilson form (which will be useful in the vicinity
of $T=T_c$; see e.g.\ Ref.\ \cite{bm79} for a derivation, and again we will set the magnetic field to zero),
\bea
S[Q]&=&\int d^dx\, \left[\frac{1}{4}\sum_{a<b} (\nabla Q_{ab})^2 + r\sum_{a<b} Q_{ab}^2 \right.\\
&&\left.{}-\frac{w}{6}\sum_{a,b,c}Q_{ab}Q_{bc}Q_{ca}-\frac{y}{12}\sum_{a,b}Q_{ab}^4 + \ldots\right]\nonumber
\eea
(we omit some further quartic terms, as well as all terms of higher order in $Q$ or $\nabla$),
and $\int {\cal D}[Q]\cdots$ denotes a functional integral over all components of $Q_{ab}(\bx)$ for all
$x$. Here $r\propto T-T_c$, and $w$ and $y$ are positive constants.
We will use the notation
\be
\langle\langle\cdots\rangle\rangle=\frac{1}{[Z^n]}\int {\cal D}[Q] e^{-S}\cdots
\ee
for averages taken using the field theory, with the $n\to0$ limit (taken at an appropriate point in the
calculation) implicit.

The analysis of the field theory begins with the identification of the appropriate saddle point value of
$Q_{ab}$, which will be position independent. This problem reduces to the same replicated action as in
the SK model, and so the appropriate saddle point has the same form: it is Parisi's hierarchical RSB
ansatz \cite{par79,mpv_book}. When we refer to RSB in this paper, in the first instance it is always this
formal scheme that we have in mind, and not a particular model (EA or SK), or a particular interpretation.
It is known that this scheme gives a formally stable solution at this mean field level, and the
propagators for the small fluctuations around it have been found. (It turns out that the saddle point
corresponds to maximizing the action \cite{mpv_book}, not minimizing it as usual. Because the solution
involves breaking
the replica symmetry, there are actually many saddle points, that are mapped to one another by symmetry
transformations. In many calculations, one should perform a sum over these distinct solutions.) More
particularly, because our interest is in the RSB theory for the EA model, in that case we mean the use of
the effective field theory with action as above, and of the RSB saddle points for that action as the
mean-field approximation.

Now we turn to the modification of this approach to study the metastate, for which the AW formulation is
the most convenient. Accordingly, we consider a system in the region $\Lambda_L$, with free or periodic
b.c.s. A subregion $\Lambda_R$ (viewed as containing both the sites in $\Lambda_R$ and the
edges that have both ends in $\Lambda_R$) will be viewed as the ``inner''
region, while the remaining sites and edges in $\Lambda_L$ constitute the ``outer'' region. When considering
correlations and overlaps, we will make use of a smaller region $\Lambda_W$, called a window; the scales
$W$, $R$, $L$ always obey the inequalities implied here (as above). We will consider copies of the
system that all experience the same disorder (random bonds) in the inner region, but some of which
experience different (independently sampled) disorder in the outer region. This enables us to consider
the correlations of spins in the window, possibly in different copies of the system, with powers of
averages over the disorder in the outer region taken first, and a final average over the disorder in the
inner region. (On occasion, independent copies of the disorder in the inner region are also used, to study
the dependence on the disorder in this region.) With this we can study the AW metastate,
because an average over the metastate means an average over the disorder in the outer region (to obtain
moments of spin correlations) for given disorder in the inner region, with a limit $L\to\infty$,
$R\to\infty$, and only finally $W\to\infty$, or else $W$ is fixed. After the metastate average(s)
has (have) been taken, one can
finally calculate moments over the disorder in the inner region to study dependence on it. Thus first
one calculates thermal averages using the Gibbs weight that depends on all the disorder. Then one
calculates averages of those over the disorder in the outer region, then over disorder in the
inner region. Finally, one takes the limit $L\to\infty$, then $R\to\infty$; whether or not the limit
$W\to\infty$ is taken depends on what one wishes to calculate. This means that the various averages in the
infinite size limit correspond with the averages in finite size (and in the two regions) through the
dictionary:
\be
\langle\cdots\rangle_\gj\leftrightarrow\langle\cdots\rangle,
\ee
where on the right the unique thermal average in finite size is meant,
\be
[\cdots]_\kj \leftrightarrow[\cdots ]_>,
\ee
and
\be
[\cdots]_{\nu(\cj)}\leftrightarrow[\cdots]_<
\ee
where $>$ and $<$ are shorthand for the outer and inner regions, respectively, in finite size.

In the replica approach, it is straightforward to allow for distinct realizations of disorder in the
outer region. We can have, say, $n_1$ replicas experiencing the first sample of disorder, $n_2$ the
second, and so on, up to $n_l$, with $n_1+n_2+\cdots n_l=n$; $n_k\to0$ for all $k=1$, \ldots, $l$ at the
end. All replicas experience the same disorder in the
inner region. After performing the disorder averages on the product of $n$ copies of the partition
functions defined in this way, the same decoupling procedure can be used, though
some attention should be paid to the inhomogeneity due to the boundary at the faces of $\Lambda_R$. Because
of the inhomogeneity, there will be terms in the effective action living on the boundary of the inner and
outer regions, however we will argue that the detailed forms of these are not required. But in
the inner region, far inside this boundary, the effective action will be the same as above, while far
outside it must consist of $l$ copies of that above, with all replica indices in the summations in the
$k$th group running from $\sum_{k'=0}^{k-1}n_{k'}+1$ to $\sum_{k'=0}^kn_{k'}$ ($n_0\equiv0$). While the
replicated effective action in
the inner region is invariant under the full symmetric (permutation) group $S_n$, that for the outer
region (and hence the action as a whole) is only invariant under the subgroup $S_{n_1}\times
S_{n_2}\times\cdots S_{n_l}$.

Next we wish to analyze this effective action, beginning by finding saddle points. The stationarity
conditions for these saddle points are local equations, and far inside the inner region are position
independent, and the same as the usual ones of RSB \cite{par79}. Far out in the outer region, the
equations for each group of $n_k$ replicas have the same position-independent form as Parisi's
\cite{par79}. There are no conditions in that region on the components of $Q_{ab}$ in which $a$, $b$ are
in different groups; those components do not occur in the effective action in that region. If we identify
$Q_{ab}$ at such a point with the average of $[\langle s_i\rangle]_{\cj_k}[\langle s_i\rangle]_{\cj_{k'}}$,
where $\cj_k$, $\cj_{k'}$ denote the different realizations of disorder in the $k$th and $k'$th copies
(which are the same in the inner region), then we expect that this is zero, just as $[\langle s_i\rangle]$
is always (in zero, and also in a weak ordering magnetic field)---that is, we expect the effect of
the common bonds in the inner region to be negligible far out in the outer region.

Consider the form of the RSB ansatz, as it is applied in the inner region. The RSB ansatz \cite{par79}
consists in dividing the matrix $Q_{ab}$ into $n^2/m_1^2$ blocks of size $m_1\times m_1$. The elements
in the blocks on the diagonal (other than the elements $a=b$) are set to one value $q_1$, and all elements
in blocks not on the diagonal are set to a (possibly different) value $q_0$. That is one step; at
each subsequent step, for example the second, the blocks on the diagonal only are subdivided further into
blocks of size $m_2$ in the same way, introducing an additional parameter $q_2$ in the diagonal subblocks,
leaving the off-diagonal blocks unchanged. This is repeated infinitely many times. Finally, as $n\to0$,
the block sizes $m_1$, $m_2$, \ldots, are assumed to go to
\be
0=n<m_1<m_2<\cdots <1.
\ee
The parameters $q_r$, $m_r$ can both be varied to maximize the effective action, however, in the limit
of infinitely many steps, the sequence becomes a function $q(x)$, $x\in [0,1]$, and this function appears
under integrals. Consequently, it does not matter if the sizes of $m_r$ are fixed (provided all differences
$m_{r+1}-m_r$ go to zero in the eventual limit) and only the $q_r$ are varied, or if $q_r$ are fixed as an
increasing sequence with a maximum at $q(1)$, and only the $m_r$ are varied. The ansatz breaks the
permutation symmetry (with group $S_n$) of the problem; consequently, performing an arbitrary simultaneous
permutations of the rows and columns of the $n\times n$ matrix in the above solution yields another equally
valid solution.

In zero magnetic field, it is found that $q(0)=0$ \cite{par79}, so we may imagine that $q_1=0$ was taken
in the
matrices. This is the same as occurs in the outer region. If we set $n_k=m_0$ (say) for all $k$, then
the off-diagonal blocks of these matrices coincide with those in the first step in the RSB in the inner
region. The block submatrices in these $m_0\times m_0$ matrices should be the same as the blocks in
the $n\times n$ matrix in the inner region, that is the matrix in the outer region has one additional step
of RSB applied before all the others. But in the limit of infinitely many RSB steps, and optimizing the
matrix sizes, one has $m_0=n_k\to 0$, and so $0=m_0<m_1$, with $m_1$ tending to zero also as the limit of
infinitely-many steps is taken. That is, after these limits, $Q_{ab}$ in the outer region is
essentially the same as in the inner region (except for effects that have measure zero in the usual
integrals over $x$ in the Parisi formalism), and we can imagine that the solution of the saddle-point
equations can be continued through the boundary between inner and outer regions to connect these solutions
in the two regions.

In the outer region, only simultaneous permutations of rows and columns
within each block of size $n_k\times n_k$ can be used to generate other valid saddle-point solutions,
corresponding to the $S_{n_1}\times S_{n_2}\times\cdots S_{n_l}$ symmetry. (There is also a symmetry under
permuting the groups of equal sizes $n_k=m_0$ themselves, however the solution is invariant under these
permutations.) That is, the outer region has imposed a symmetry-breaking field on the system in the inner
region, that acts through the boundary of the latter. The remaining saddle-point solutions obtained by more
general permutations (members of $S_n$) that were valid in the inner region (in isolation) presumably must
not be used because they lead to a free energy penalty at the boundary of the two regions. We note that,
in particular, using the saddle points we have proposed to calculate the free energy of the (say) first
group gives the same answer as usual, as it must because the other copies that see different disorder drop
out as $n_k\to0$ for $k\neq 1$. Moreover, use of any other solutions (if they exist) would imply that the
replica symmetry is broken even further; thus our ansatz corresponds to assuming that the replica symmetry
is not broken even further by the boundary between inner and outer regions.

Now we use this ansatz to obtain information about the AW metastate of the RSB theory. Throughout,
we will illustrate the ideas using low-order moments, though distribution functions can also be
obtained. First we will consider moments in which averages over the inner and outer regions (or over the
metastate $\kj$ and the bond distribution $\nj$) are performed simultaneously. The simplest is for the
average overlap, defined by (and taking the limit)
\be
\left[\left[\langle
s_i\rangle_\gj^2\right]_\kj\right]_{\nu(\cj)}=\lim \left[\left[\langle s_i\rangle^2\right]_>\right]_<
\ee
for $\bx_i$ in the window (the position does not matter, by translation invariance of the average).
In replica language, the average on the right can be obtained by taking two replicas $a\neq b$ in the
same group, so they experience the same disorder in the outer region. Then the result is the same as in
the RSB literature \cite{dy,par83}. The value of the bilinear spin expectation can be represented by
the saddle point
for $Q_{ab}$. Due to summation over the different saddle points, or equivalently by summing over
distinct $a$ and $b$ in $1$, \ldots, $n_1$, the result is (in the limit $n_1\to0$)
\bea
\frac{1}{n_1(n_1-1)}\sum_{a\neq b} Q_{ab}&=&\int_0^1 dx\, q(x)\\
&=&\int_0^1 dq\, q P(q)
\eea
where again by definition $P(q)=dx/dq$ (and $P(q)=0$ for $q>q_{\rm EA}$). In the infinite-size point of
view, each state $\gj$ in the
metastate can be analyzed into ($\cj$-dependent) pure states $\Gamma_{\cj,\alpha}$ labeled by $\alpha$
(again, these are well defined because we are considering infinite size) as in eq.\ (\ref{eq:puredecomp}),
and then the above calculation gives the disorder and metastate average
$\left[[\cdots]_\kj\right]_{\nu(\cj)}$ of the weighted sum of overlaps, that is of
\be
\sum_{\alpha,\beta} w_{\gj,\alpha} w_{\gj,\beta}\, q_{\alpha,\beta},
\ee
where the overlap $q_{\alpha\beta}$ for two pure states labeled $\alpha$, $\beta$
was defined in eq.\ (\ref{overlaps}). Note that the same notation for pure states
$\langle\cdots\rangle_\alpha$ can be used here as in the decomposition of $\rj$ earlier, because all of
them belong to the set of pure states of the infinite-size system with bonds $\cj$, and they do not depend
on the state $\gj$ or $\rj$ in whose decomposition they appear, though the weights ($w_{\gj,\alpha}$ or
$\mu_{\cj,\alpha}$) do.

More generally, the disorder and metastate average $\left[\left[P_{\cj,\gj}(q)\right]_\kj
\right]_{\nu(\cj)}$ of the overlap distribution (\ref{eq:ovdist}) can be
obtained by studying higher moments, that is, higher powers of $q$ under the $\left[ [\cdots]_\kj
\right]_{\nu(\cj)}$
average, or using a moment generating function. This still involves the use of just two
copies of the system, or in replica terms, two values $a$, $b$ from the same group. The calculation can
be carried out exactly as in Ref.\ \cite{par83}, even though here we take the per-spin overlap in the
window $\Lambda_W$ only. In the replica field theory, there will be correlations between components of
$Q_{ab}$ at distinct points in
the window, but these can be neglected to leading order as $W\to\infty$, so the $Q_{ab}$ field is simply
replaced by its saddle-point value in each place. The result is then that the averaged distribution is
$\left[\left[P_{\cj,\gj}(q)\right]_\kj\right]_{\nu(\cj)}=P(q)$ as defined above, and as in Ref.\
\cite{par83} up
to the differences of definition as noted already, and the use here of well-defined pure states.

By this point it should be clear that when averages over the disorder and the metastate (or over disorder
in the inner and outer regions in finite size) are performed {\em together}, the results take the same
form as in the old RSB literature, up to i) the use of the window in place of the whole
system when calculating overlaps, ii) the dropping
of subleading terms due to correlations of the $Q$ field variable, and iii) the use of well-defined
infinite-size pure states. Naturally then, the same also applies to what was called ``non-self-averaging''
\cite{ybm,mpstv}, and to ultrametricity \cite{mpstv}. As an example of the former, one can show that
\bea
\label{eq:non_self}
\lefteqn{W^{-2d}\sum_{i,j\in\Lambda_W}\left[\left[\langle s_i\rangle_\gj^2\langle
s_j\rangle_\gj^2\right]_\kj
\right]_{\nu(\cj)} \neq}&&\\
&&\qquad\qquad W^{-2d}\sum_{i,j\in\Lambda_W}\left[\left[\langle
s_i\rangle_\gj^2\right]_\kj\right]_{\nu(\cj)}\left[\left[\langle
s_j\rangle_\gj^2\right]_\kj\right]_{\nu(\cj)}\non
\eea
even as $W\to\infty$, unless $q(x)$ takes the same value at almost every $x$, which is not true in the
RSB scheme anywhere below $T_c$, except at $T=0$. To derive the
statement, note that in the replica approach as formulated here, the left hand side involves four copies
of the system, or choosing four distinct replicas from the group of size $n_1$, while the right-hand
side involves choosing only two, but then squaring the result, effectively dropping some restrictions.
The inequality of the two sides is connected with the breaking of replica symmetry. More generally, for
the distribution function one has, for example \cite{mpstv},
\bea
\lefteqn{\left[\left[P_{\cj,\gj}(q)P_{\cj,\gj}(q')\right]_\kj\right]_{\nu(\cj)}=}&&\nonumber\\
&&\qquad\frac{2}{3}\left[\left[P_{\cj,\gj}(q)\right]_\kj\right]_{\nu(\cj)} \
\left[\left[P_{\cj,\gj}(q')\right]_\kj\right]_{\nu(\cj)} \nonumber\\
&&\qquad{}+
\frac{1}{3}\left[\left[P_{\cj,\gj}(q)\right]_\kj\right]_{\nu(\cj)}\delta(q-q')
\eea
as $W\to\infty$, and so for general $\cj$
\be
P_{\cj,\gj}(q) \neq P(q)
\ee
even at leading order as $W\to\infty$ (except at $T=0$). ($P_{\cj,\gj}(q)$ consists of a countable number of
$\delta$-function spikes that depend on $\cj$, whereas $P(q)$ is a smooth function, except for a
$\delta$-function at $q=q_{\rm EA}$. Actually, it is interesting to ask whether the limit $W\to\infty$ of
$P_{\cj,\gj}(q)$ even exists, and if so, why. According to the RSB theory, it does, and we return to this
in Sec.\ \ref{sec:rigid}.) Ultrametricity in the organization
of the pure states \cite{mpstv} works out similarly. It can be interpreted as the ultrametric structure
as $W\to\infty$ of the per-spin window overlaps $q_{\alpha\beta}$
of the pairs $\alpha$, $\beta$ occurring in the decomposition of any {\em one} state $\gj$ drawn from
the metastate (here ``any'' means, more strictly, with probability one with respect to the product
distribution $\nu(\cj)\kj$).

The question of non-self-averaging of the overlap distribution was an early target in NS, see Ref.\
\cite{ns96a}. They showed that in the EA model, within rigorous constructions of the thermodynamic limit
and of an overlap distribution, non-self-averaging could not occur because of translation ergodicity of
certain distributions. It was understood later that both of the two constructions of the limit given
there involve an average over the metastate \cite{ns96b}. Therefore, it is of interest to study whether
the non-self-averaging seen above should be attributed to the randomness due to the metastate $\kj$ or
to the disorder $\nj$. For this we consider averages that distinguish inner and outer regions, for example,
\be
W^{-2d}\sum_{i,j\in\Lambda_W}\left[\left[\langle s_i\rangle_\gj^2\right]_\kj\left[\langle
s_j\rangle_\gj^2\right]_\kj
\right]_\nj.
\ee
In this case, to leading order as $W\to\infty$, we use the saddle-point values of $Q_{ab}Q_{cd}$ where $a$,
$b$ are from one group, say the first with $n_1$ members, and $c$, $d$ are from another, say the second,
with $n_2$ members. Because only permutations within each group must be summed over, one finds
to leading order that the average is equal to
\be
\left(\int_0^1 dx\, q(x)\right)^2,
\ee
which should be contrasted with eq.\ (\ref{eq:non_self}) above, indicating that non-self-averaging above
can be blamed on the dispersal of the metastate.
We can again introduce an overlap distribution, this time the metastate-averaged one, defined by
\be
P_\cj(q)=\left[P_{\cj,\gj}(q)\right]_\kj,
\ee
which can be calculated under the $\nj$ average by finding its moments. From the preceding calculation,
the fluctuations of this distribution appear likely to be benign in the $W\to\infty$ limit, and this
turns out to be the case. By modifying the calculations of Ref.\ \cite{mpstv} for this case, one finds
that the disorder-induced fluctuations of $P_\cj(q)$ vanish within the usual approximation of dropping
correlations between distinct points, and so
\be
P_\cj(q)=P(q)
\ee
[i.e.\ its own $\nj$-average] to leading order as $W\to\infty$. The lack of dependence on $\cj$ agrees
with the results of NS. Thus the so-called non-self-averaging in RSB is due to the dispersal of the
metastate, as NS \cite{ns96b} argued had to be the case if RSB is to be valid. This means that RSB
agrees with the non-standard picture.

We can go a step further in the direction of averaging before squaring correlations. If we recall the
definition of the MAS $\rj$ [see in particular eq.\ (\ref{rjavdef})], we see that averaging using the
metastate before squaring a spin-correlation function can give moments of averages in $\rj$. (This
then corresponds to the other construction of a thermodynamic limit in Ref.\ \cite{ns96a}, which was for
the state itself.) The one-point function $\left[\left[\langle s_i\rangle_\gj\right]_\kj^2\right]_\nj$
(and other odd-spin correlations) will vanish in the present case of zero magnetic field, though
not in the case of non-zero magnetic field; this average corresponds to $\langle\langle
Q_{ab}\rangle\rangle$ with $a$, $b$ from different groups, and so the general result is $q(0)$. As in the
discussion in Sec.\ \ref{sec:frac}, if this differs from $\int_0^1 dq\,q(x)$ then it implies that the
metastate is supported on more than one Gibbs state; the converse is
also clear (assuming that the Gibbs states are incongruent \cite{fh1}). Thus we can say quite generally,
independent of the details of the RSB scheme (and hence even if fluctuations play an important role),
that as long as Parisi's function $q(x)$ characterizes the RSB, $q(0)\neq \int dx\,q(x)$ if and only if
the metastate is nontrivial. Note that this occurs within
RSB at $T=0$, where the picture can be called chaotic pairs instead of non-standard SK, because as $T\to0$,
$q(x)\to 1$ for $x\neq0$, while $q(0)=0$ for all $T$. By contrast, the earlier results of this section,
which involved only integrals over $x$, could not distinguish RSB from restoration of replica symmetry
at $T=0$. The fact that $q(0)$ differs from $\lim_{x\to0}q(x)$ at $T=0$ makes a difference.

Returning to the effective field theory and the case of zero magnetic field, correlations of an even
number of spins such as
\be
\left[\left[\langle s_i s_j\rangle_\gj\right]_\kj^2\right]_\nj
\ee
will not vanish. This 2-point function is the same as the correlation function $C(\bx)$ defined above in
eq.\ (\ref{cx}). In the replica formalism we require
\be
C(\bx)=\langle\langle Q_{ab}(\bx)Q_{ab}({\bf 0})\rangle\rangle,
\ee
where here $a$ and $b$ are from {\em different} groups. The expectation value of this component of
$Q_{ab}$ is zero (corresponding to the vanishing of the $1$-point function as mentioned), so this time
we must consider the correlations in the $Q$ variables; this correlation function of $Q_{ab}$ is a
connected one. These have been studied in Ref.\ \cite{dkt}, and see Ref.\ \cite{ddg_book} for a review.
The result is
\be
C(\bx)\sim \frac{1}{|\bx|^{d-\zeta}},
\ee
(up to a constant factor) as $|\bx|\to\infty$, where
\be
\zeta=4.
\ee
(For the corresponding connected correlation function in the presence of a magnetic field, below the
de Almeida-Thouless line, $\zeta'=3$ \cite{ddg_book}.)
The application of this correlation function to obtain properties of the metastate is one of the
central results of this paper. It was obtained for fluctuations around the mean-field
theory; hence this result should be correct for $d\geq 6$. The result was obtained at nonzero temperature
(in fact, not far below $T_c$), but we expect that it is universal and holds at all $T<T_c$, including
at $T=0$. The consequences of this result for the variance of the disorder average
$\left[P_{\cj,\rj}(q)\right]_\nj$ of the per-spin window-overlap distribution $P_{\cj,\rj}(q)$ defined
in Sec.\ \ref{sec:frac} were already discussed there. They can be interpreted as saying that the dimension
of the space of pure states (at zero temperature, alternatively as the logarithm of the number of
ground states) that can be distinguished inside a window of size $W$ scales as $W^{d-4}$ for $d\geq 6$.

We can obtain further information about the overlap distribution $P_{\cj,\rj}(q)$ by examining its higher
moments in the same way. For leading order results, we can always make use of clustering which holds in
pure states. Then the disorder averages of moments of $q$ with respect to $P_{\cj,\rj}(q)$ can be obtained
from $k$-point functions
\be
\left[\left[\langle s_{i_1} s_{i_2}\cdots s_{i_k}\rangle_\gj\right]_\kj^2\right]_\nj
\ee
by averaging over the positions $\bx_{i_l}$ over the window $\Lambda_W$ ($l=1$, \ldots, $k$). In the
replica formalism this becomes
\be
\langle\langle Q_{ab}(\bx_{i_1})\cdots Q_{ab}(\bx_{i_k})\rangle\rangle.
\ee
It vanishes for $k$ odd, while for $k$ even, if we use only the part of the effective action expanded to
quadratic order in $Q_{ab}$ (i.e., neglecting interactions between the modes) we obtain
\be
\prod_{l=1}^{k/2}\langle\langle Q_{ab}(\bx_{i_{2l-1}})Q_{ab}(\bx_{i_{2l}})\rangle\rangle
+\hbox{permutations},
\ee
using Wick's theorem. On averaging the positions over the window, we find that the distribution $\left[
P_{\cj,\rj}(q)\right]_\nj$ is a Gaussian with variance $\sim W^{-(d-\zeta)}$. Note that, to make the
statement that the distribution approaches a Gaussian as $W\to\infty$ precise, we must first define a
variable $\tilde{q}=qW^{(d-\zeta)/2}$, and then the distribution for $\tilde{q}$ approaches a limit as
$W\to\infty$, and that limit is Gaussian with variance of order 1. We will leave this implicit in what
follows.

In order to justify the assumptions that led to the Gaussian distribution, we should examine the role of
the interactions among the modes $Q_{ab}$ (for {\em all} $a$, $b$) in calculating the $k$-point functions.
Because of the ``massless'' or power-law propagators (2-point functions of $Q$), this proceeds similarly as
in critical phenomena. For massless modes with a gradient-squared term in the interaction and when the
lowest order interactions are cubic, the interactions can be usually be neglected (they are irrelevant)
for $d>6$ as they lead to subleading corrections if included using perturbation theory. In the present
case, the interaction terms themselves are not singular at $\bk\to0$, and have no singular dependence on
the replica indices. But the $2$-point function of direct interest has $a$ and $b$ in distinct groups and
diverges as $|\bk|^{-4}$ as $\bk\to0$. These facts could mean that the interactions are not irrelevant for
all dimensions $d$ above $6$, but only for $d$ above a lower bound that is larger than $6$. We conclude
that the Gaussian form of the distribution will be valid (within perturbation theory about the RSB saddle
points) in sufficiently high dimensions. The study of detailed correction terms to correlation functions,
which could establish the precise dimension above which the Gaussian behavior is valid, is beyond the scope
of this paper.

In a similar fashion as for $P_{\cj,\gj}(q)$, one can study higher moments of the distribution
$P_{\cj,\rj}(q)$ itself, to see if it is equal to its disorder average at leading order as $W\to\infty$.
For $\left[P_{\cj,\rj}(q)P_{\cj,\rj}(q')\right]_\nj$, for example, this involves four copies or replicas
from distinct groups, and similarly for higher moments. If interactions among the modes are neglected, the
use of Wick's theorem leaves some terms that contain the $2$-point function $\langle\langle
Q_{ab}Q_{cd}\rangle\rangle$, with all of $a$, $b$, $c$, $d$ in distinct groups. Results from Refs.\
\cite{dkt,ddg_book} imply that this is zero. This and a similar argument as before about interactions
between the modes being irrelevant leads to the conclusion that $P_{\cj,\rj}(q)$ is equal to its disorder
average (using $\nj$) in sufficiently high dimensions, in leading order as $W\to\infty$.

If one does not introduce the scaled variable $\tilde{q}$, then the distribution for $q$ tends to a
$\delta$-function in the limit $W\to\infty$, and this distribution is independent of $\cj$. This agrees
with the results of Refs.\ \cite{par_comm,mari} and with the result of NS that it must be self-averaged
\cite{ns96a,ns96b}. The fact that the scaled distribution is Gaussian and self-averages is a stronger
statement than these.

Similarly as in studying the $\cj$-dependence of $P_{\cj,\rj}(q)$, we can also consider higher moments of
the $2$-point correlation function in $\rj$, as we did in the fractal-cluster models; for the $2k$th
moment, we again use $2k$ copies or distinct replicas in the replica field theory. We find
\be
\left[\left[\langle s_i s_j\rangle_\gj\right]_\kj^{2k}\right]_\nj \sim (2k-1)!!\, C(\bx_i-\bx_j)^k
\ee
at large separation, where the implicit coefficient is independent of $k$.
Here we again used the non-interacting approximation for the modes, which is valid for the large
$|\bx_i-\bx_j|$ asymptotics in sufficiently high dimensions. For each $k$, the function behaves as
$\sim |\bx_i-\bx_j|^{-k(d-\zeta)}$
with $\zeta=4$ as before. This is very different from the fractal-cluster model, in which it $\sim
|\bx_i-\bx_j|^{-(d-\zeta)}$ for all $k$. The result, in particular the $k$-dependence, shows that,
at any fixed well-separated positions $i$ and $j$ and at leading order, the $2$-point function
$\left[\langle s_i s_j\rangle_\gj\right]_\kj$ is Gaussian distributed with mean zero. We expect this
to hold even at zero temperature.

The results we have obtained show that RSB theory predicts that the metastate for the EA spin glass behaves
according to the so-called non-standard picture. More than that, we have obtained quantitative information
about the metastate: not only is the total number of pure states in the metastate uncountable,
but its size can be quantified using the logarithm of the number of ground states (or at nonzero
temperature the dimension of the manifold of states) that can be distinguished in a window of size $W$,
which grows as $W^{d-\zeta}$ with $\zeta=4$.

Some of the results of this section (those concerning the agreement with the results of Ref.\ \cite{ns96a})
were obtained in Ref.\ \cite{mari}. That paper takes a different point of view from the present one on
several issues, especially about the use of pure states in finite versus infinite size \cite{ns02}.
When discussing the metastate briefly, those authors stated that the results {\em disagreed} with the
non-standard picture
as well as with the standard one. They seem to identify the non-standard picture with an ansatz about
finite-size approximants (a topic we discuss in Sec.\ \ref{sec:finite} below) from Ref.\ \cite{ns96b},
and state that it is ``not
a surprise'' that the non-standard picture is inconsistent in that case, in apparent acceptance of the
arguments in Ref.\ \cite{ns98}. We discuss the arguments of the latter paper in
Sec.\ \ref{sec:critique}.

\subsection{Summary and further scenarios}

At this stage it may be helpful to apply the basic results from the replica analysis to general scenarios.
The analysis holds independent of most details of the field theory, but we will continue to assume that
Parisi's function $q(x)$ describes the pattern of any RSB. Then the non-standard picture provides the
``master recipe'' for constructions of metastates, from which various other scenarios can be obtained
as limiting cases. In general, the assumptions are that $q(x)$ is an integrable function (in the
Lebesgue sense, in general) and that it is monotonically increasing ($q(x)\geq0$ due to the ordering
field, and $q(x)\leq 1$ from its definition).

As in the old (standard) interpretation of RSB, in all cases, if $q(0)\neq q(1)$, so that replica
symmetry is broken, then there are many pure states that are thermodynamically relevant, not only a
pair related by a spin flip. Further, if $q(1)\neq \int_0^1 dx\,q(x)$, then a typical state $\gj$ drawn
from the
metastate is non-trivial; its decomposition involves many pure states, with self-overlaps equal to
$q(1)=q_{\rm EA}$. Also, if the breakpoint $x_1<1$, so that there is a plateau, with $q(x)=q(1)$
for $x\in(x_1,1]$, then the pure states in the decomposition of $\gj$ are countable. In addition, we
have demonstrated that if $q(0)\neq \int_0^1 dx\,q(x)$, then the metastate is non-trivial, as shown
by the two-point correlations $C(\bx)$ in the MAS $\rj$ decaying to a value smaller than in a typical
$\gj$.

These rules of interpretation lead to various possible scenarios, every one of which is consistent with
the requirements of translation invariance producing self averaging \cite{ns96a} (c.f.\ Sec.\
\ref{sec:scen}). The cases in which either the metastate, or a Gibbs state drawn from it, are trivial,
are uniquely determined up to one or two parameters. These are (i)
the scaling/droplet picture, with $q(x)=q_{\rm EA}$ for all $x\in [0,1]$; (ii) the chaotic pairs picture
\cite{ns96b},
with $q(x)=q(1)$ for $x>0$, but $q(0)\neq q(1)$, as we explained; (iii) a variant picture in the same class
as the old standard picture: it has a trivial metastate, which implies that $q(x)=q(0)$ for $x<1$,
but $q(1)\neq q(0)$. Thus in this variant (mentioned in Ref.\ \cite{ns96b}) the unique Gibbs state $\gj$
in the support of the metastate
is nontrivial, and contains uncountably many pure states. Its MAS $\rj$ is the same as itself, and so
$\gj$ resembles $\rj$ in our earlier discussion, and presumably should have power-law
correlations as there. It has ultrametric structure trivially, and no non-self-averaging or CSD. The
remaining possibilities, in which (iv) state and metastate are both non-trivial, can have forms of $q(x)$
that differ from that in the RSB mean-field theory, including cases in which it is not continuous.
A particular class of variants (also in Ref.\ \cite{ns96b}) is those in which $q(x)$ is not constant
for $x\in(0,1)$, but has no plateau at $q_{\rm EA}$; it need not be continuous at $x=1$. In this case,
all features of the previous RSB picture are preserved (including so-called non-self-averaging and
non-trivial ultrametric structure of $\gj$), except that the number of pure states involved in any typical
$\gj$ is uncountable.

\section{Critique of arguments against the non-standard picture}
\label{sec:critique}

We have argued that the structure resulting from the RSB theory for the EA model is the non-standard
SK picture, advanced as a logical possibility by NS \cite{ns96b}. However, these authors also argued,
though as they said, not completely rigorously, that the non-standard picture can be
ruled out; the main argument in this direction appears to be that in Ref.\ \cite{ns98}
(to be referred to as NS98), but the conclusion has been repeated in several others.
Because the non-standard picture is what emerges from RSB theory, here we will carefully
re-examine the arguments of Ref.\ \cite{ns98}. We will find that, not only is the argument not fully
rigorous, it appears to have very little force. First, we will recapitulate the argument briefly.

The argument, as it appears in Sec.\ IV of NS98, has three main parts, and a conclusion. The first part is
an argument for the ``invariance of the metastate''. The latter is the
following statement: in the definition of the NS metastate probabilities as the asymptotic frequency of
each (in general, mixed) state in the sequence of systems of size $L$ as $L\to\infty$ (as discussed in
Sec.\ \ref{sec:csd} above), a b.c.\ on the hypercube of size $L$ must be given for each
$L$. The invariance of the metastate is the property that the same metastate is obtained from any two
sequences of
b.c.s that are related for each and every $L$  by a flip on the boundary spins or bonds
at that $L$. The flips that can be used are completely arbitrary, both in their position dependence at
each $L$, and their $L$ dependence. This property is easily seen to hold if the definition of the
AW metastate average as an average over the bonds in the outer region is used, because the flips can be
absorbed into the disorder average by change of variables (a typical gauge-invariance argument in
spin-glass theory, which holds because there is no applied magnetic field). Assuming the equality of the
AW and NS metastates, the invariance property as stated follows.

The second part of the argument is a claim that the invariance of the metastate should imply that the
distribution is in some sense ``uniform'' on the states on which it is supported. It is admitted that
the notion of uniformity is not clearly defined, which prevents the argument from being entirely rigorous.
The strongly-disordered model is invoked as an example in which the metastate is known to be uniform in
a precise sense.

The third part of the argument is the statement that a change in a bond at finite distance from the origin
produces a calculable change in the relative thermal weights (at non-zero temperature) of each pure state
in any state drawn from the metastate, or likewise in the weight of any pure state in the MAS. (The former
is called the ``strong covariance property''.) It is stated that these results can be
shown rigorously.

Finally, it is argued that the third point is inconsistent with the claimed
``uniformity'' of the metastate, which would not permit such changes in weight. It is only here that the
nature of the non-standard picture enters, as it assumes that each state in the metastate is a mixture
of states not all of which can be related by a global spin flip, and so the argument does not apply to
the chaotic pairs picture.

The first and third parts of the argument seem strong enough, and we will make no objection to them.
The problem lies with the second part, and the claim of a ``uniform'' distribution over the states.
Before starting our critique of that part, we point out that in order to obtain the conflict claimed
at the conclusion of the argument, the uniformity that is supposed to be required in the second part of
the argument must be {\em strictly} uniform, not just approximately. This is because if the distribution
is only required to be approximately uniform, then this may be compatible with the dependence on the
disorder $\cj$ at finite distance as required in the third part.

We will examine the arguments for the second part in more detail. (Some of the initial arguments of NS98
are actually stated for the chaotic pairs picture, but the distinction appears to make little difference
to this part of the argument, because
for one thing, chaotic pairs and non-standard SK are the same at zero temperature anyway.) The first
reason given, after discussion of the strongly-disordered model
and the observation that such a uniform distribution would rule out the chaotic pairs picture with a
countable infinity of states, is that it would be ``unreasonable'' for, say, a periodic b.c.\ in every size
to produce a given state with a frequency of say $0.39$, and for also the antiperiodic
b.c.\ in every size to produce that state with the same frequency (and likewise for many
other choices of boundary condition). But this is not clear at all. First, the situation should not be
confused with saying that, in one particular size---say, a very large one--the periodic b.c.\ produces the given state and so does the antiperiodic, and all other, b.c.s. That
would mean the state is insensitive to b.c.s, and is not expected when there is CSD.
However, the stated situation is in fact about the {\em distribution} of states (and NS98 is clear
about this). The same
state should be produced with probability $0.39$ for antiperiodic as well as for periodic conditions, but
that does not mean it must occur in the same sizes for both conditions (or in the same samples, in the
disorder-averaging definition of the metastate). One can expect that the bonds just inside the boundary,
which change randomly as the size is increased because they are among the ones added, can compensate the
effect of the b.c.\ change. Thus, the layer of edges normal to a face of the $d$ dimensional
hypercube of size $L$ contains almost as many possible sign choices as the boundary bonds or spins do, and
may be able to simulate approximately the effect of the original b.c.\ in some other
samples after a change in boundary condition. The same holds for thicker layers of bonds as well.
This type of behavior may be implied by the equivalence of the two definitions of the metastate, also,
and should be the reason for the invariance of the metastate.

We are willing to accept for the sake of argument that some sort of uniformity is
implied by the invariance of the metastate, but it is an effectively uniform distribution over
b.c.s (the next remarks in NS98 also refer to such a choice). We will now examine the
consequences of this. Heuristically, we can then think of the following model for such a metastate:
in a given size $L$ we have a probability distribution over all configurations of the spins on the
boundary of the hypercube, with equal probability for each. Each of these configurations will be viewed
as a fixed-spin b.c.\ on the thermal (Gibbs) state of the spins in the interior, and the
bonds in the interior will all be viewed as being at finite distance from the origin (that is, $R=L$
here). Then we examine the state in a window of size $W$ ($W\ll L$). Finally, we can let $L$ and $W$ go
to infinity, maintaining $W\ll L$. (NS98 here consider the MAS, though this may not be necessary.)
NS98 state that ``it seems unreasonable that this last \ldots metastate \ldots can have anything other
than a uniform distribution over the pure states''.

This sounds like an error. Let us again consider the case of
zero temperature for simplicity; then the state produced in the limit for each b.c.\ is a
spin configuration that minimizes the energy subject to the boundary condition. So each b.c.\ maps
to a state in the window. The minimization is a complicated non-linear problem, and
depends on all the bonds $J_{ij}$ in the system of size $L$. Now finding the probability distribution
for the states in the window is rather like the following: Suppose one has a single real variable $x$
taking values in a finite interval, and a function $f$ on that interval, and that we
write $y=f(x)$. If $x$ has uniform probability density on the finite interval, then the probability density
for $y$ is not uniform, but is proportional to $|dx/dy|$, the Jacobian of the transformation from
$x$ to $y$, and in general is not uniform. But NS's remark sounds like omitting the Jacobian factor.

In fact, even though, in the model we are considering, each boundary spin configuration has equal
weight, these weights are of order $2^{-{\cal O}(L^{d-1})}$ because of the invariance that models
invariance of the metastate. There
must be many distinct b.c.s (boundary spin configurations) that correspond to the same
state (spin configuration) in the window, because there are ${\cal O}(L^{d-1})$ spins on the boundary,
but the logarithm of the number of  states distinguishable in the window is only at most
${\cal O}(W^{d-1})$ (indeed, only ${\cal O}(W^{d-\zeta})$ according to the arguments in this paper).
But (except in some exceptional circumstances
we will describe in a moment) there is no reason at all why the states in the window must
all have the same total weight (probability in the metastate). It is most likely that a ground state is
selected by a given b.c.\ by a ``majority vote'' involving the boundary spins and the bonds
$J_{ij}$ in the interior. The total weight of a given ground state is proportional to the number of
different b.c.s that produce it, which will depend on many details. Furthermore, the
weights depend on the bonds $J_{ij}$ through the minimization process, so that if some finite number of
bonds (say, inside the window) are changed, each ground state can change. For a large enough window, such
a change (presumably) will not render two initially-distinct ground states the same, nor change the
weight of each. Similarly, for the mixed state in the positive temperature case, states can change, but also
the relative weights of distinct pure states in the mixed state for some (collection of) boundary
condition(s) can change, consistent with the strong covariance property.

It should be clear that the situation is very different in the case of the strongly-disordered model. There
(in any finite size) the spins on each tree in the minimum spanning forest have their relative values
determined by the signs of the bonds on the edges of the tree, and the overall sign is determined by the
spin at the unique vertex at which the growing tree first touches the boundary (a change in that spin
produces a flip of all spins on that tree), independently of all the other boundary spins. Thus the
uniform distribution on the boundary spins produces a uniform distribution over the possible flips of
the trees that intersect a given window, and thus over the distinct ground state visible in the window.
But this is very different from the case of the EA spin glass, in which, even if there were clusters of
spins (determined once and for all by $\cj$) that all flip together (independently of others), whether
the flip occurs depends on the outcome of a competition between many boundary spins that encounter the
cluster and the bonds within the cluster.

We conclude that the second part of the NS98 argument---that invariance of the metastate leads to a
uniform distribution over Gibbs states in the metastate (or over pure states in the MAS) of the EA spin
glass---has little force: not only because the notion of uniformity has not been defined, but because
there is no reason for uniformity of {\em these} distributions at all. At best, the argument as a whole
boils down to a quantitative question of whether the approximate uniformity that {\em might} be required
by the invariance of the metastate (and is anyway plausible) is compatible with the strong covariance
property that requires some dependence on the disorder $\cj$. At present, there appears to be no such
quantitative analysis that could answer this question in the negative and thus rule out the non-standard
picture.

\section{Discussion}
\label{sec:disc}

In this section we discuss some remaining open questions, and give answers to some of them. In some cases,
these answers are more speculative than the content of the rest of the paper. As in Sec.\
\ref{sec:rsb_meta}, in this section we will proceed by using the RSB effective field theory, and see what
it predicts, without attempting to prove or disprove these predictions.

\subsection{Lower dimensions}

We have discussed the spin-glass state of the EA model according to RSB theory at the mean-field level only.
Especially in view of the power-law correlations in the phase (and in the MAS), it is natural to use a
point of view based on critical phenomena and the renormalization group and apply it to the effective field
theory. Then we would expect that our conclusions, in particular the value of the exponent $\zeta=4$,
would be valid even beyond mean-field theory at least in sufficiently high dimensions. For the critical
behavior of the spin glass, the upper
critical dimension, above which mean-field exponents are valid, is six (certain additional effects enter
between six and eight dimensions \cite{gmb,fs}). For the spin-glass phase below $T_c$
(or below the de Almeida-Thouless line), the situation is less clear, because of the technical difficulties
of calculations. Nonetheless it is expected \cite{dkt,ddg_book} (from work that did not relate the relevant
correlation exponent to the metastate) that the result $\zeta=4$ also holds down to six dimensions, but
may be modified for $d<6$.

We can formulate a scenario for lower dimensions that is motivated by the behavior of the
strongly-disordered model \cite{ns94,jr}. We know that $\zeta\leq d$, so if it reaches $\zeta(d)=d$
a change in the metastate
can be expected, and not all features of the mean-field RSB metastate will hold in lower $d$; this
criterion also appeared previously \cite{dkt,ddg_book}. Generally speaking, the values of exponents
will change continuously as the dimension $d$ is lowered, unless a fixed point becomes unstable when
$d$ passes below some value, in which case a jump in exponents could occur.  Thus a natural
scenario is to expect no breakdown of the RSB picture at six dimensions. Then one would expect that
for $d<6$, $\zeta$ will move continuously away from $4$. We would not expect the picture of the
spin-glass phase necessarily to break down at four dimensions either, if $\zeta$ tends to increase with
decreasing $d$. Even the sign of this derivative seems not to be known definitely (assuming the
low-temperature fixed point is not unstable), but there is recent numerical work (again, with a different
definition of $\zeta$) that suggests that $d-\zeta\simeq 1$ in $d=4$ \cite{npr}, and earlier that
$d-\zeta\simeq0.5$ in $d=3$ \cite{mari,janus}. Then the critical dimension
above which there are many pure states in the metastate
would seem to be close to the lower critical dimension for the spin-glass phase (below which there should
be no transition at non-zero temperature), which is believed to be larger than two and less than or equal
to three. It is tempting to expect that these two critical dimensions may coincide, even though it is not
obvious that they must. We note that there is a recent rigorous proof that the two-dimensional EA model
has only a single pair of infinite-size ground states in zero magnetic field, but the proof is for the
{\em half-}plane \cite{adns}; from a rigorous point of view, the case of the plane remains open. If the
result holds for the plane, then it is consistent with the scenario that there are many states
(i.e.\ a non-trivial metastate) only above the lower critical dimension.

It has been suggested \cite{ddg_book} that for $d\leq 6$, $\zeta(d)$ in the low-temperature phase
($T<T_c$) is related to the value of the exponent $\eta(d)$ of the {\em critical} theory ($T=T_c)$,
by
\be
\zeta(d)= \frac{d+2-\eta(d)}{2}.
\ee
Whether this can be supported by analysis of the effective field theory seems unclear,
but it is a possibility within the same scenario.

However, there are serious questions about the correctness of the preceding scenario. There is apparently
no consensus on whether a change of stability of the perturbative low-temperature fixed point that
controls the low-temperature spin-glass phase occurs or not just below six dimensions. Worse, there is a
clear problem with the existence of the de Almeida-Thouless line for $d\leq6$; early work was unable to
identify a corresponding stable perturbative fixed point \cite{br}. The result has been confirmed
\cite{ptd} and no good solution to the problem has been proposed. This calls into question the existence
of a transition in a magnetic field, and with it the occurrence of RSB even at zero field, for $d<6$
(and perhaps even for $d>6$, depending on whether the coupling constants in Ref.\ \cite{br} lie in the
basin of attraction of the origin \cite{mb}). Further, recent work that studied $d\to6$ from above
\cite{mb} found indications of a loss of stability of the RSB solution below $d=6$, namely the
disappearance of the de Almeida-Thouless line, and the vanishing of the break point $x_1$ in $q(x)$,
as $d\to6^+$. (However, as we will discuss in a moment, the latter does not necessarily mean full
restoration of replica symmetry.) The interpretation of those results has been disputed \cite{pt}, and
see also Ref.\ \cite{temes13} for recent work. In any case, if RSB disappears below $d=6$, even if
the spin-glass phase does not disappear at $h=0$, then this scenario differs from the previous one,
in that the logarithm of the number of states $\sim W^{d-\zeta}$ does not become of order one at $d=6$,
even though RSB disappears.

The scaling/droplet theory was originally advanced \cite{bm1,macm,fh} as a low-dimensional picture that
might explain the
lower critical dimension. A typical approach to understanding the behavior at the lower critical dimension
is to consider the ground state and its low-energy excitations, as in the scaling/droplet approach. As
we discussed, the RSB effective theory approach is a typical approach that comes from high dimensions. The
possibility that (if the lower critical dimension is lower than that below which there is only a single
state in the metastate) the scaling/droplet picture applies for intermediate dimensions cannot be ruled out
using the RSB effective field theory; neither, in that case, can the possibility of the chaotic pairs
picture \cite{ns96b} for intermediate dimensions be ruled out. Chaotic pairs would emerge from the RSB
field theory if $x_1$, the breakpoint in $q(x)$, goes to zero for $T<T_c$ and $d$ less than some critical
value (as suggested for $d<6$ by the results of Ref.\ \cite{mb}), and if $q(x)$ has to be viewed as
discontinuous at $0$; more generally, if $q(x)\to q_{\rm EA}$ for all $x\neq0$, and $q(0)=0$.
(Indeed, the latter does occur as $T\to0$ in the Parisi solution, and the metastate
we have found at $T=0$ could also be termed chaotic pairs instead of non-standard.)
As we explained in Sec.\ \ref{sec:rsb_meta}, this would imply a non-trivial metastate and CSD.
(It is not clear what would happen to $\zeta$ in this scenario.)
However, according to other authors \cite{pt,temes13}, this scenario does not occur in the RSB field theory
close to, and just below, six dimensions. We note that if chaotic pairs applies in non-zero magnetic
field (``chaotic singles''), then $0\neq q(0)\neq q(1)$, and a transition below which replica symmetry
is broken must occur on some de Almeida-Thouless line in the $T$-$h$ plane.

Which of these latter possibilities would occur in intermediate dimensions depends on the mechanism
by which full RSB---that is, with $q(x)$ not constant for $x\in(0,1]$---is destroyed. In the first of
our earlier scenarios for the behavior in lower dimensions, the logarithm of the number of states went
to order one (and most likely to zero) at the lower critical dimension for full RSB, suggesting that in
lower dimensions scaling/droplet behavior takes over. In the second scenario, the logarithm of the number
of states could jump to zero, or the system could go over to the chaotic pairs picture, which still has
many states, and only the properties relating to non-constant $q(x)$ for $x\neq0$ would be destroyed.

\subsection{Rigidity in RSB}
\label{sec:rigid}

While discussing window overlaps $q_{\alpha\beta}$ of pure states $\alpha$ and $\beta$ in Sec.\
\ref{sec:frac}, (and see also Sec.\ \ref{sec:rsb_meta}),
we mentioned that it is not obvious from first principles that they are independent of the size and
location of the window used. The spin expectations $\langle s_i\rangle_\alpha$ and $\langle
s_i\rangle_\beta$ might vary randomly and independently with position, and then the overlaps would
fluctuate, with the per-spin overlaps approaching zero. Indeed, within the results of the RSB field
theory, this is exactly what we found for the overlap of two typical pure states drawn from the
decomposition of the MAS $\rj$: the per-spin overlap in a window fluctuates around zero, with width
tending to zero, and by a similar analysis one can see that the average using $\mu_{\cj,\alpha}
\mu_{\cj,\beta}$ of the correlation between overlaps in two distant windows tends to zero with
separation, just as $C(\bx_i-\bx_j)$ does.

In the case of the MAS, the number of pure states involved is uncountable, and this behavior is probably
not surprising. However, for two pure states $\alpha$, $\beta$ involved in the decomposition of any one
$\gj$ drawn from the metastate, the RSB theory does predict that the per-spin overlap approaches a constant
as $W$, the size of the window, tends to infinity, and the constant is independent of the position of
the center of the window. This follows from, for example, the expected form of the two-point function
\bea
\lefteqn{\left[\left[\langle s_i s_j\rangle_\gj^2\right]_\kj\right]_\nj}&&\nonumber\\
&\sim&\left[\left[\sum_{\alpha,\beta} w_{\gj,\alpha}w_{\gj,\beta}
\langle s_i\rangle_\alpha\langle s_i\rangle_\beta\langle s_j\rangle_\alpha\langle
s_j\rangle_\beta\right]_\kj\right]_\nj\\
&\sim&
\int dq\, P(q) q^2
\eea
as $|\bx_i-\bx_j|\to\infty$, by using clustering in both the pure states and the RSB field theory. This
shows that, according to the RSB theory, the pure states $\alpha$, $\beta$ ``know'' that they should have
the same value of the per-spin overlap in each region of space, up to local statistical fluctuations
that average out if we average over two large but well-separated windows. This is what allows the
overlap distribution $P_{\cj,\gj}(q)$ for given $\cj$ and $\gj$ to have a non-zero limit as $W\to\infty$
(and in that limit it does not self-average).

This property might be somewhat surprising, even though as we saw, the overlaps are non-zero only for pure
states from the same Gibbs state $\gj$.  It can be viewed as a kind of ``stiffness'' or ``generalized
rigidity'' in the RSB theory, related to the gradient-squared term in the action of the effective theory.
That allows the two-point function of the field $Q_{ab}$ to tend
to a non-zero tensor in each replica-symmetry-broken ordered state (each of which corresponds to a
single saddle point of the effective action). That is, without the stiffness, the two-point function
would tend to zero with separation. The preceding remarks show that the spin-glass ordered state possesses
a high degree of rigidity according to the RSB theory. (The influence of the distant outer region on the
inner region, which enforces essentially vanishing overlap for a pair of pure states drawn from $\rj$
was also due to the same stiffness.) This may be contrasted with the relative softness
that emerges when an EA spin glass is characterized in terms of the exponent $\theta$ \cite{fh}
(or $y$ \cite{bm1}) that describes the scaling of the free energy cost for forcing in a domain wall
by a change of boundary condition, and which obeys the inequality $\theta\leq (d-1)/2$ \cite{fh};
in high dimensions this makes it much smaller than what is usual for a domain wall, as in a ferromagnet,
where the value is $d-1$, or the analogous property for a small twist in a system with a continuous
broken symmetry, where one finds $d-2$. In the RSB theory, the generalized rigidity or stiffness is
connected with the breaking of replica symmetry, similar to the general situation for broken symmetries
\cite{anderson}. It appears to be in some sense dual to the softness mentioned, as it affects different
correlations. By analogy with ordinary broken symmetries, one might expect the generalized rigidity to go
continuously to zero at the lower critical dimension for RSB, whatever dimension that may be.

\subsection{Metastate interpretation and finite-size systems}
\label{sec:finite}

We have emphasized that interpreting RSB theory in the language of pure states requires that the
infinite-size limit be taken. But in contrast, historically most discussion of spin glasses has
been in terms of finite-size systems. The question of finite-size systems remains important, for example
because they are the only ones that can be studied numerically. This raises the question of
what exactly the RSB theory and the metastate approach can say about finite-size systems.

In relation to the use of the RSB mean-field theory, or other approaches that yield the same results,
it must be emphasized that mean-field theory is intrinsically a method for infinite size. The assumption
of some mean or molecular field that has to be determined self-consistently, and which may depend on a
choice of a state (in a system with given $\cj$) makes sense in infinite size, as the molecular field at
a site depends on neighboring ones, which in turn depend on further neighbors, and so on out to infinity.
(For spin glasses one may think of the TAP/cavity approach here.) Thus underlying such an approach there
is a natural connection between solutions of the self-consistency equations and pure states in the
``official'' definition for infinite size in a short-range system. (For the SK model, there is a
similar situation involving different notions \cite{pan_book} in place of the DLR Gibbs states.)
Mean-field theory is
therefore not intrinsically good at capturing finite-size effects; it has to be supplemented by a
field-theory method that can incorporate fluctuations around the mean field. It would be interesting to
study this within the RSB theory; for some recent work in this direction see Ref.\ \cite{cpv}. But
occasional claims in the literature (e.g\ Ref.\ \cite{mari}) that the RSB mean-field theory is directly,
and only, concerned with the behavior in finite-size systems should be viewed as suspect. When pure
states are invoked, these involve the use of infinite size, otherwise they are not well defined.

Turning to the metastate, it is constructed from a limit of finite-size systems. The existence of the limit
implies, and in fact requires, that the finite-size system is approximated by the infinite-size object
arbitrarily closely as the size increases---albeit in terms of probability distributions for a random
object. In either the AW or NS constructions of the metastate, the state $\Gamma_{\cj,L}$ in a system of
size $L$ is approximately given by a state $\gj$ drawn from the metastate $\kj$, in the sense that
the marginal joint distributions for the set of spins in a window $\Lambda_W$ for $\Gamma_{\cj,L}$ and
$\gj$ agree approximately; here $W$ must be somewhat smaller than $L$ (as before). The last restriction is
crucial (this has been emphasized in NS's papers, and also in Ref.\ \cite{mari}.) As both objects being
compared are random even when $\cj$ is given (say in the intermediate size window $\Lambda_R$), so that we
do not know which $\Gamma_\cj$ is the ``right'' one for our $\Gamma_{\cj,L}$, it will be necessary to
gather statistics about the marginal distribution for the set of spins in $\Lambda_W$ (say as the bonds
in the outer region vary), and compare the distributions of these conditioned marginals; then one is
comparing approximate partial versions of $\kj$. At present, rather little is known about the rate of
convergence of such approximations as $L\to\infty$.

For finite-size systems, an ``empirical'' window overlap distribution can be defined directly, without first
assuming a decomposition into pure states; it is a proxy for the marginal joint distribution for all the
spins in the window. In this, we use two configurations $S$, $S'$ drawn independently from the same
Gibbs state $\Gamma_{\cj,L}$, and calculate their overlap, say in a window,
\be
\widehat{q}=W^{-d}\sum_{i\in \Lambda_W} s_is_i'.
\ee
The distribution
\be
P_{\cj,L}(q)=\sum_{S,S'}\Gamma_{\cj,L}(S)\Gamma_{\cj,L}(S') \delta(q-\widehat{q})
\ee
(where $\Gamma_{\cj,L}(S)$ is viewed as the Gibbs weight for configuration $S$) of this for given $\cj$
is an empirical distribution
that can be compared with a corresponding window overlap $P_{\cj,\gj}(q)$ that may be obtained from the
state $\gj$ drawn the metastate. In the infinite-size system, the distribution $P_{\cj,\gj}(q)$ consists of
$\delta$-function peaks, while that in the empirical distribution $P_{\cj,L}(\widehat{q})$ is found to
contain broadened peaks (which in fact contain many $\delta$-functions, corresponding to the individual
spin configuration overlaps). To study the AW metastate, which may be the more convenient choice,
the statistics of the empirical window overlap distribution should be studied as the bonds in the
outer region are sampled from $\nj$, keeping those in the inner region fixed.

In practice, most studies
use finite-size overlaps defined globally, with $\Lambda_W$ in the definition of
$\widehat{q}$ replaced by $\Lambda_L$, the whole system (in particular, this empirical overlap distribution
was studied in the recent works, Refs.\ \cite{ykm,bill,middleton,bill2}). While such global overlaps may
be misleading in some systems \cite{fh1}, it is not clear if that is the case for the EA model, and
some studies that compared the distributions of global and window overlaps found that they were similar
\cite{mprr}. In any case, such overlap distributions show clear signs that they depend on the disorder
(i.e.\ $P_{\cj,\gj}(q)$ is ``non-self-averaging''). To study the metastate, it is necessary to disentangle
the effect of the distant disorder in the outer region from that of the disorder in the inner region, and
for that the window overlap distribution must be used.

However, the most direct way to obtain $\zeta$ is probably to construct numerically an approximate version
of the MAS $\rj$. To study the overlap distribution in $P_{\cj,\rj}(q)$, it may not be necessary to average
both distributions $\Gamma_{\cj,L}(S)$ over $J_{ij}$s in the outer region. Instead one can choose one,
presumably typical, set of bonds using $\nj$. Then one can perform a Monte Carlo simulation that runs
simultaneously on that system and on a second system that has the same bonds in the inner region, but an
independently-chosen sample of bonds in the outer region, and calculate the empirical window overlap
between the two systems. Then one can repeat this with the first system the same, but the second having
a different independently-chosen set of bonds in the outer region. This can be repeated many times to
obtain an average over the bonds in the outer region of the second system. An average over the bonds in
the outer region of the first system could be performed also, but should not be necessary as it was
presumably typical; thus an average over a small number of independent copies (with the bonds in the
inner region fixed) should be sufficient. For $W$, $R$, and $L$ large enough, the resulting finite-size
approximation to the overlap distribution $P_{\cj,\rj}(q)$ should display the behavior described above,
if RSB theory is correct. One may also try an average over the bonds in the inner region. Another
approach would be to look directly for dependence of the ground state in the window on the bonds in the
outer region, for well-separated scales $W$, $R$, $L$. In either case it may be difficult to obtain an
accurate value of $\zeta$ because of the need for well-separated scales $W$, $R$, $L$, however, the
qualitative behavior of the different overlap distributions
$P_{\cj,\rj}(q)$ and $[P_{\cj,\gj}(q)]_\kj$ (both of which are asymptotically independent of $\cj$ as
$W\to\infty$ \cite{ns96a}; the latter equals the Parisi $P(q)$ in that limit) should be strikingly
different if RSB theory is correct. In fact, these distributions could be studied even for
{\em small} $W$, and could exhibit clear signs of CSD; such calculations should be easier than measuring
$\zeta$. Even a window so small that it contains just two spins could be fruitful.

As we emphasized above, the infinite-size Gibbs states, and the metastate, are defined in terms of spin
correlations (and hence overlaps) in windows whose size is held fixed as the system size goes to infinity.
Thus the theory developed here does not directly apply to the properties of global overlaps, or other
global constructs, in large finite systems. A different theory would be required in order to characterize
the infinite-size limit of such global objects. From the RSB point of view, the starting point is again
the effective field theory, now with (say) periodic b.c.s, and no outer region. From this
point of view, the behavior of global overlaps would indeed be expected to be similar to that of the
window overlaps. However, full access to the AW metastate is not available in this approach, due
to the absence of the outer region. We can instead refer to the approach leading to the NS metastate.
If we fix a finite size, and study moments under the distribution of random bonds, then this corresponds
to drawing a sample from the NS metastate (again, neglecting the distinction between window and global
correlations and overlaps), and the final average over bonds (and then taking the
$L\to\infty$ limit if desired) implies an average over the metastate as well. That is, all averages
correspond to the type $[[\cdots ]_\kj]_\nj$ in the work above, whose behavior is known from the old
replica calculations (with non-self-averaging, and so on). This contrasts with the first construction of
an infinite-size overlap distribution in Ref.\ \cite{ns96a}, in which the infinite-size limit was taken
before any $\nj$ average; in that case, the distribution must self-average \cite{ns96a}. These points
may give some hints about the form of a theory for the global overlaps.

\section{Conclusion}
\label{sec:conc}

The old De Dominicis-Young-Parisi interpretation \cite{dy,par83} of the formal RSB mean-field theory
for Ising
spin glasses is still frequently invoked (though perhaps the infinite-size limit is not mentioned as often
as it used to be). In this paper, we have used the replica theory itself to derive directly its
probabilistic interpretation for the EA model.
The main parts of the approach and results are (i) when speaking of pure states, we used only the concept of
pure states in an infinite system, which is well-defined; (ii) we applied the RSB effective field
theory approach to characterize the metastate of the theory---the metastate \cite{ns96b}, which is a
probability distribution over Gibbs states, accounts for CSD, and is again a well-defined notion in an
infinite system. The metastate was found to be non-trivial and to be in accord with the so-called
non-standard picture, advanced as a logical possibility by NS \cite{ns96b} to agree with rigorous
results \cite{ns96a,ns96b} as well as with RSB (note that some of the results, pointing to agreement
with rigorous results, appeared previously in Ref.\ \cite{par_comm,mari});
(iii) the picture was extended with the quantitative result that an exponent
$\zeta$, which can be viewed as an effective fractal dimension for clusters of spins, takes the value
$\zeta=4$ when the dimension of space is high; (iv) arguments \cite{ns98} that were thought almost
to rule out the non-standard picture were examined and found lacking. These results mean that a
consistent interpretation of the RSB theory for short-range spin glasses has been obtained within the
theory itself. Further, a strikingly simple statement can be made more generally: $q(0)\neq \int_0^1
dx\,q(x)$ only if the metastate is nontrivial (CSD occurs). In addition to these results, we proved
a theorem that describes the metastate-average state of the infinite-range SK model exactly, providing a
counterpoint to the results for the short-range model.

The ``non-standard'' interpretation due to NS \cite{ns96b} becomes much more detailed as a result of
using results from RSB to characterize it; for example, in addition to finding the value of $\zeta$,
its distribution functions can be studied. \
The name ``non-standard'' is unfortunate, because this
picture should be the canonical one for the RSB theory. We propose to call the particular metastate that
has emerged simply the ``RSB metastate'', and call this approach to RSB theory the ``metastate
interpretation'' (in more recent work, e.g.\ Ref.\ \cite{ns07}, NS refer to it as the ``RSB picture'').
In our view, any description of the physics of RSB in terms of many pure states, and
so on, that does not include the fact of CSD and the need to use the metastate is deficient.

We emphasize that a key property derived here from the RSB theory is that the number of
pure states that exist in the EA model (in infinite size) is uncountable (it was proved that it must be
whenever the non-standard picture holds in Ref.\ \cite{ns07}). However, when attention is
restricted to a window of size $W$, the logarithm of the number that can be distinguished within that
volume grows only as $W^{d-\zeta}$ and $\zeta\geq 1$,
that is, its entropy is subextensive. This number differs from what has been stated in some other places in
the literature. For example, the logarithm of the number of solutions of the TAP equations was found to
be proportional to $N$, the system size \cite{te,dggo,bm_tap}, though it is not clear if all of them
should be viewed as corresponding to pure states (however, this number does agree with the result for the
SK model). On the other hand, the number of pure states occurring
in the decomposition of a single pure state $\gj$ must be countable, because of the plateau in $q(x)$
within the usual RSB theory; hence it will not
grow exponentially with the size of a window. We should mention that
while the number of states in infinite size is uncountable, the theory predicts that not all of these
states can be seen in a single sample of given finite size, even as ``finite-size pure states'', which
is a somewhat ill-defined concept, and even allowing for the restriction to a window of size $W$ less
than the system size $L$. In such a sample, only some of those in the countable set for a single
$\gj$ can be seen. The larger number can be seen only by allowing the bonds in the outer region to vary,
or by considering a range of different sizes. Returning to the case of infinite size, it is also
interesting to note that the growth found here of the logarithm of the number, $\sim W^{d-\zeta}$, in the
EA model violates a proposed possible upper bound
\cite{ns01} by the corresponding logarithm of the number of ground states, $\sim W^{d-6}$ \cite{jr},
in the strongly-disordered model \cite{ns94}, at least for all $d\geq 6$, where $\zeta=4$.

While we don't claim to have established either the existence of many pure states in the EA model, or
that the RSB metastate is the correct one, the consistency of the interpretation suggests that both
statements could be true when a spin-glass phase exists (i.e.\ for $d\geq 3$), up to caveats such as that
the overlap distribution in the MAS that was found here to be Gaussian in high dimensions could be modified
in lower dimensions.
In trying to distinguish the RSB theory from alternatives such as the scaling/droplet theory, numerical
work should examine whether there is CSD, or alternatively strong dependence of the state on the bonds
in the outer region. Either of these is closely connected with the existence of many pure states
\cite{ns92}.

We remark that similar results should be expected for the EA model in a magnetic field, a
problem we did not consider much, though it seems that then $\zeta$ is replaced by $\zeta'=3$ \cite{dkt},
and for other models, such as with Potts, XY, or Heisenberg spins, and also for power-law
interactions.

We close with a wish list of topics for future research. The thermodynamic results of the RSB theory
can be recovered from the cavity method without using replicas \cite{mpv1}; it would be nice to
have also a non-replica derivation of the exponent $\zeta$ in the short-range models, even a heuristic one,
as this could illuminate the origin of this power law. Other desirable results, if the RSB theory is to
be established, include a rigorous proof in sufficiently high dimensions either that there are many pure
states, or that there is a transition in non-zero magnetic field (since for both, the scaling/droplet
picture assumes/predicts not). Further, a theory for dimensions near
the lower critical dimension that makes predictions about the scaling and the number of pure/ground
states there, but {\em allowing} for the existence of many pure states, would complement the RSB
mean-field and effective field theory approach that comes down from high dimensions. Again, even a
heuristic theory could provide insight into the disappearance of many pure states at the
lower critical dimension for RSB.

\acknowledgments

The author is grateful to the physicists from whom he learned about spin glasses over many years,
in particular
A.P. Young, D. Sherrington, P. Goldbart, A. Houghton, A.J. Bray, M.A. Moore, D.S. Fisher, D.A. Huse,  C.M.
Newman, and D.L. Stein, and especially to Newman and Stein for answering questions about their work, and
for comments on an earlier version of the manuscript. He thanks S. Tatikonda, I. Gruzberg, M. Madiman,
and M.A. Moore for useful discussions, in particular Tatikonda and Madiman for raising the subject of
exchangeability and for helpful references, and Moore for discussions of dimensions close to six. He
also thanks Stein
and Gruzberg for encouragement to write the paper. This work was supported by NSF grant no.\ DMR-1005895.

\appendix

\section{AW metastate and MAS of the SK model}
\label{sec:aw-sk}

For completeness and for illustration/comparison with the EA model, we address here first the analog of
the AW metastate for the SK model, at zero temperature. The results can be obtained from Ref.\ \cite{ns03}
(especially the proof of Theorem 3). Then we state and prove a theorem about the MAS (using the AW
metastate) in the infinite-size limit of the SK model, which extends the preceding result. None of the
arguments involves replicas.

In this Appendix, we will use the SK Hamiltonian in the form
\be
H= -\frac{1}{\sqrt{N}}\sum_{(i,j)}J_{ij}s_is_j
\ee
for $N$ spins; the edges connect all pairs of spins, and we assume independent Gaussian distributions
for each $J_{ij}$, with variance $J_0^2$ independent of $N$. We will consider a subset containing $M$
of the spins as
constituting the window or inner region, and the remaining $N-M$ spins as the exterior. NS showed that,
as $N\to\infty$ in a given sample, each spin configuration in the window is a ground state infinitely
often \cite{ns03}. We will first look for the probability distribution for the (finite-size) ground-state
spin configuration, conditioned on fixed bonds $J_{ij}$ for $i$ and $j$ both in the window; thus the
probability distribution is induced from the disorder in the remaining bonds, those outside the window,
and also those that connect the window to the exterior. This is the analog of the AW metastate, although
we note that the definition of infinite-size Gibbs states in Sec.\ \ref{sec:gibbs} does not apply in the
SK model.

The result is largely a simple symmetry (gauge invariance) argument. First we notice that, as $N\to\infty$,
the bonds within the window contribute to the Hamiltonian by terms of order
$1/\sqrt{N}\to0$, while their number is fixed, so they can be neglected. (A rigorous justification for
this statement in NS \cite{ns03} uses the limit $N\to\infty$, $T\to0$ with $T\sqrt{N}\to\infty$.)
Then because the magnetic field
is zero, the probability distribution for the spin configuration in the window is invariant under a flip
$s_{i_0}\to-s_{i_0}$, $J_{i_0 j}\to -J_{i_0j}$ for any one site $i_0$ in the window, and for all $j$ not
in the window, and everything else fixed. Hence given a ground state that occurs with
some probability, there is equal probability for the ground state with $s_{i_0}$ reversed. We conclude
that in the AW meta\-state as $N\to\infty$, each of the $2^M$ spin configurations in the window occurs
with equal probability. More precisely, ground states actually occur in pairs related by a flip, and
the $2^{M-1}$ possible ground state pairs occur with equal probability.

The question of the analog of the NS metastate in the SK model remains open. It is however clear
\cite{ns03} that (as expected \cite{ns97}), if for given $\cj$ one considers a sparse (and
$\cj$-independent) sequence of sizes $N_0$, $N_1$, \ldots, say with $N_{k+1}=1000 N_k$ for all $k$, the
choice of ground state in the window is dominated by a set of bonds that is independent of those at
smaller sizes, and then by the law of large numbers the frequencies with which the ground state pairs
appear in the window (along the sparse sequence) will approach their probabilities in the AW metastate,
namely uniform over the $2^{M-1}$ pairs; this gives the NS result \cite{ns03}. It would be interesting
to prove the existence of the NS metastate itself as a limit, that is to prove that the frequencies
with which ground-state pairs occur as $N$ increases through {\em all} the integers, for given $\cj$,
converge to the frequencies in the AW metastate. It might be possible to do this heuristically by using
a version of the cavity method \cite{mpv,mpv_book}; it would be necessary to study the correlations
between the ground states (in the window) at different $N$, as $N$ increases.

Now we turn to a theorem about the MAS constructed using the average over the AW metastate. This can be
viewed as generalizing the preceding result for the zero-temperature AW metastate to non-zero temperature.
That is because at zero temperature, a state drawn from the metastate is a ground state. The metastate is a
probability distribution on the ground states, and if viewed in terms of spin configurations it thus
becomes a probability distribution on spin configurations. That is, at zero temperature the MAS and the
metastate itself are equivalent (either with or without the presence of spin-flip symmetry or magnetic
field).

We state the {\bf Theorem}: the infinite-size MAS (using the AW metastate) at zero magnetic field and
all $T$, $0\leq T\leq \infty$, is the product of independent uniform distributions on the Ising spins
$\{\pm 1\}$ in any window of $M$ vertices, independent of the bonds $J_{ij}$ with both ends in the window.
That is, each of the $2^M$ spin configurations in the window has equal probability, regardless of the
$J_{ij}$s in the window.

We note that the statement is true trivially in the high temperature limit, and should follow from the
cavity method for all $T>T_c$ (where the molecular fields vanish). At zero temperature, it is
equivalent to what was shown above, as already mentioned. In general, it is plausible because the effect
of the couplings in the window on the spins should be negligible in the limit, and because it resembles
what we found in the EA model, keeping in mind that the SK model has no useful notion of distance and
every spin in the window is on its boundary, so that the power-law correlations don't exist, and in the
fractal-cluster point of view each cluster consists of a single spin. If we compare with the EA model
at fixed $d$, then $W^{d-\zeta}=W^d=M$ and so in a sense we have $\zeta=0$ in the SK model. However, this
may not be the best way to define $\zeta$ for the SK model. If instead we write $W^{d-\zeta}=
M^{1-\zeta/d}$ and assume the SK model corresponds to $d\to\infty$, then it agrees with the theorem, but no
information about $\zeta$ in the SK model is obtained.

We now sketch a proof (which is essentially rigorous) of the theorem using results from probability
theory. We begin with
the joint probability distribution of the bonds and spins in a system of $N$ spins, which is $\nu(\cj)
e^{-H(S)/T}/Z$ (for $T=0$ the distribution is defined as the limit $T\to0^+$; the following argument
holds for all $T$ including $T=0$.)
If a limit of this as $N\to\infty$ exists, then it is defined in terms of its marginal distributions for
$s_i$ and $J_{ij}$ with $i$, $j$ in subsets of finite size $M$, which are exactly the distributions of
interest. From the metastate point of view, because the AW metastate average is defined as an average
over the bonds in the outer region (and taking the limit), this construction gives the MAS \cite{aw}
(it is similar to the second construction in Ref.\ \cite{ns96a}). It is not necessarily clear that a
(unique) limit actually exists. But as before, by a ``standard compactness argument'' (see e.g.\ the
appendix of Ref.\ \cite{ns97} for a similar argument, and Appendix A14 of Ref.\ \cite{billingsley1} for
the diagonal method used to complete it),
it can be proved that there are subsequences of the sequence $N=1$, $2$, \ldots, along which the
distribution converges (weakly) to a limit \cite{billingsley1,billingsley2}.

Next we will consider the behavior of any one of these limits. A key property that we will use is called
``exchangeability'' in
the probability literature. Exchangeability of a probability distribution means that it
is invariant under the action of the permutation (symmetric) group. In our case, a permutation
$i\to\sigma(i)$ has the effect $s_i\to s_{\sigma(i)}$, $J_{ij}\to J_{\sigma(i)\sigma(j)}$. The SK
Hamiltonian for finite $N$ is obviously invariant under such a transformation, and hence so are
the partition function and the joint probability distribution of $S$ and $\cj$ for finite $N$. In the case
of a probability distribution for an infinite set of variables, we require only that it be invariant under
finite permutations that move only a finite number of vertices. (More precisely, in the literature
distributions of symmetric matrices with this invariance property are termed ``weakly'' exchangeable, and
also matrices are termed ``arrays''.)

The $J_{ij}$ form an off-diagonal symmetric matrix, and we can place the variables $s_i$ in the diagonal
positions to form a symmetric matrix $(s_i,J_{ij})$. For an infinite symmetric matrix whose distribution
is exchangeable, there is a characterization of the probability distribution due (independently) to Aldous
and to Hoover \cite{ald_hoov} (see also Ref.\ \cite{pan_book}), which is a generalization of the classical
de Finetti theorem (see e.g.\ Refs.\ \cite{billingsley1,pan_book,df}) that applies to infinite exchangeable
{\em sequences} (vectors). The
Aldous-Hoover representation theorem for weakly-exchangeable symmetric arrays says that there exist
functions $g$, $f$, with two and four arguments respectively, and where $f$ is invariant under exchange of
the middle two arguments, such that the random variables $(s_i,J_{ij})$ are equal ``in distribution'' to
functions $s_i=g(u,u_i)$, $J_{ij}=f(u,u_i,u_j,u_{ij})$ of the set of random variables $u$, $u_i$, $u_{ij}$
(as $i$, $j$ range over the positive integers, and $i\neq j$), all of which are independent and uniformly
distributed over the interval $[0,1]$, and $u_{ij}=u_{ji}$ for all $i$, $j$. (See e.g.\ pages 22--28 of
Ref.\ \cite{pan_book} for a proof.) Some idea of the meaning of
this can be obtained from the simpler statement of de Finetti for an infinite exchangeable sequence,
say $s_i$, which again can always be represented in the same way as $s_i=g(u,u_i)$ (in this case no
$u_{ij}$s enter). If $g$ were independent of $u$ (as a function), this would say simply that the $s_i$
are independent and identically distributed. A non-trivial dependence on $u$ means that each spin
is independent if $u$ is fixed (i.e.\ if we condition on $u$), but the distribution depends on $u$.
As $s_i=\pm 1$, this can be described by saying that the spins all experience the same magnetic field, but
it is a random variable. It should be clear that this is exactly the situation one finds in the
infinite-range model of a ferromagnet, for which mean-field theory is exact. There $u$ is essentially
the molecular field, which is the same on every site; for $T<T_c$ it is random but takes one of only two
values.

As the subsequence $N\to\infty$ limit of the joint distribution of $(s_i,J_{ij})$ in the SK model should
again be weakly exchangeable in the above sense, we can apply the Aldous-Hoover theorem to it. In the
resulting representation for $(s_i,J_{ij})$, we can now eliminate some of the dependencies on the
independent variables. First, the $J_{ij}$ are known to be independent and identically distributed
[under the marginal distribution, which is $\nu(\cj)$]. If the $J_{ij}$s depended on the global and
vertex variables $u$, $u_i$, $u_j$, that would produce correlations between the $J_{ij}$s (when they share
a vertex, in the case of the $u_i$s). This implies that $J_{ij}$ can be represented as a function of a
bond variable $u_{ij}$, and does not depend on the global and vertex variables $u$, $u_i$, $u_j$. (The
function itself is determined also, because the marginal distribution for $J_{ij}$ is Gaussian.) Likewise,
for $s_i$, as in the preceding discussion of the de Finetti theorem, if it depended on the value of the
global variable $u$, there would be ferromagnetic correlations between $s_i$ and $s_j$. That is clearly
not the case in the SK model at zero magnetic field. So $s_i$ is effectively a function of a random $u_i$
only, and in fact $s_i$ takes the values $\pm 1$ with equal probability, by symmetry. Hence the $s_i$'s and
$J_{ij}$'s for $i$, $j$ in the finite subset (window) $\{1,2,\ldots,M\}$ are all independent. This shows
that the joint distribution of $(s_i,J_{ij})$ in the window is independent of the subsequence limit used,
which establishes that the $N\to\infty$ limit of the joint distribution exists, without the
need to use subsequences. Finally, the MAS is obtained by conditioning on the $J_{ij}$ in the window
(or on those in a larger ``inner'' region, but this clearly makes no difference).

Note that this provides an alternative route to the
NS result above, one that works in the case $T\to0$ at finite $N$, followed by the limit $N\to\infty$.

\end{document}